\documentclass[aps,prl,preprint,superscriptaddress]{revtex4-2}

\usepackage{graphicx}
\usepackage{bm,enumerate,amsmath,amssymb,amsthm}

\newcommand{\twocol}{0}
\newcommand{\Vsi}{$\mathrm{V_{Si}}$}
\newcommand{\micron}{\,$\mathrm{\mu m}$}
\newcommand{\micronsq}{\,$\mathrm{\mu m^2}$}
\newcommand{\microsec}{\,$\mathrm{\mu s}$}

\newcommand{\pqubit}{$+$ qubit}
\newcommand{\mqubit}{$-$ qubit}

\newcommand{\pmqubits}{$\pm$ qubits}

\newcommand{\ket}[1]{|#1\rangle}

\begin{document}


\title{Quantum sensing with duplex qubits of silicon vacancy centers in SiC at room temperature}


\author{Kosuke Tahara}
\email[]{ktaha@mosk.tytlabs.co.jp}
\author{Shin-ichi Tamura}
\author{Haruko Toyama}
\author{Jotaro J. Nakane}
\author{Katsuhiro Kutsuki}
\affiliation{Toyota Central R\&D Labs., Inc., Nagakute, Aichi 480-1192, Japan}

\author{Yuichi Yamazaki}
\affiliation{National Institutes for Quantum Science and Technology, Takasaki, Gunma 370-1292, Japan}
\author{Takeshi Ohshima}
\affiliation{National Institutes for Quantum Science and Technology, Takasaki, Gunma 370-1292, Japan}
\affiliation{Department of Materials Science, Tohoku University, Aoba, Sendai, Miyagi 980-8579, Japan}


\date{\today}

\begin{abstract}
The silicon vacancy center in Silicon Carbide (SiC) provides an optically addressable qubit at room temperature
in its spin-$\frac{3}{2}$ electronic state.
However, optical spin initialization and readout are less efficient compared to those of spin-1 systems,
such as nitrogen-vacancy centers in diamond, under non-resonant optical excitation.
Spin-dependent fluorescence exhibits contrast only between $\ket{m=\pm 3/2}$ and $\ket{m=\pm 1/2}$ states,
and optical pumping does not create a population difference between $\ket{+1/2}$ and $\ket{-1/2}$ states.
Thus, operating one qubit (e.g., $\left\{\ket{+3/2}, \ket{+1/2} \right\}$ states) leaves the population in the remaining state ($\ket{-1/2}$) unaffected,
contributing to background in optical readout.
To mitigate this problem, we propose a sensing scheme based on
duplex qubit operation in the quartet,
using microwave pulses with two resonant frequencies to simultaneously operate $\left\{ \ket{+3/2}, \ket{+1/2} \right\}$ and $\left\{ \ket{-1/2}, \ket{-3/2} \right\}$.
Experimental results demonstrate that this approach doubles signal contrast in optical readout and
improves sensitivity in AC magnetometry compared to simplex operation.
\end{abstract}


\maketitle

\section{Introduction}
Spin qubits in color centers exhibit considerable quantum-sensing capabilities
because of their unique combination of high spatial resolution, sensitivity, and operability under ambient conditions\cite{balasubramanianNanoscaleImagingMagnetometry2008,mazeNanoscaleMagneticSensing2008,rondinMagnetometryNitrogenvacancyDefects2014,degenQuantumSensing2017,barrySensitivityOptimizationNVdiamond2020}.
The signal contrast of spin-dependent fluorescence is a critical parameter for optimizing sensitivity.
Sensitivity scales linearly with contrast,
whereas other parameters, such as spin coherence time or fluorescence intensity,
show square-root scaling\cite{phamEnhancedMetrologyUsing2012,rondinMagnetometryNitrogenvacancyDefects2014,barrySensitivityOptimizationNVdiamond2020}.

Although various applications have been demonstrated using nitrogen-vacancy (NV) centers in diamond\cite{wrachtrupProcessingQuantumInformation2006,balasubramanianNanoscaleImagingMagnetometry2008,mazeNanoscaleMagneticSensing2008,rondinMagnetometryNitrogenvacancyDefects2014},
color centers in Silicon Carbide (SiC) are attractive because of mature material and device technologies.
High-quality SiC wafers and power electronic devices are commercially available and are used in industrial and consumer products.
Several color centers in SiC are room-temperature spin qubits, i.e.,
optically detected magnetic resonance (ODMR) is observable at room temperature and coherent manipulation of the spin state is possible
\cite{koehlRoomTemperatureCoherent2011,widmannCoherentControlSingle2015,wangCoherentControlNitrogenVacancy2020,sonDevelopingSiliconCarbide2020,castellettoQuantumSystemsSilicon2023}.
Among them, the silicon vacancy (\Vsi) center\cite{widmannCoherentControlSingle2015,carterSpinCoherenceEcho2015,tarasenkoSpinOpticalProperties2018} is one of the most established options
because its structure and physics are understood, and it can be created reproducibly in a controlled manner.
In 4H-SiC, the negatively charged \Vsi\ at the cubic ($k$) site, the V2 center, functions as a room-temperature qubit\cite{ivadyIdentificationSivacancyRelated2017}.
This specific color center is denoted as \Vsi\ in this paper.
Notably, the spin-optical dynamics of \Vsi\ have been studied comprehensively\cite{liuSiliconVacancycenters2024}.
Moreover, in terms of materials engineering, \Vsi\ can be reproducibly created in SiC by irradiations with electrons, neutrons, or
ions\cite{krausThreeDimensionalProtonBeam2017,ohshimaCreationSiliconVacancy2018,kasperInfluenceIrradiationDefect2020}.
Furthermore, the affinity for electron device technology has already been demonstrated through
electrical detection of spin states\cite{niethammerCoherentElectricalReadout2019} and device thermometry\cite{hoangThermometricQuantumSensor2021}.

\begin{figure}
  \if\twocol1
    \includegraphics[width=8.5cm]{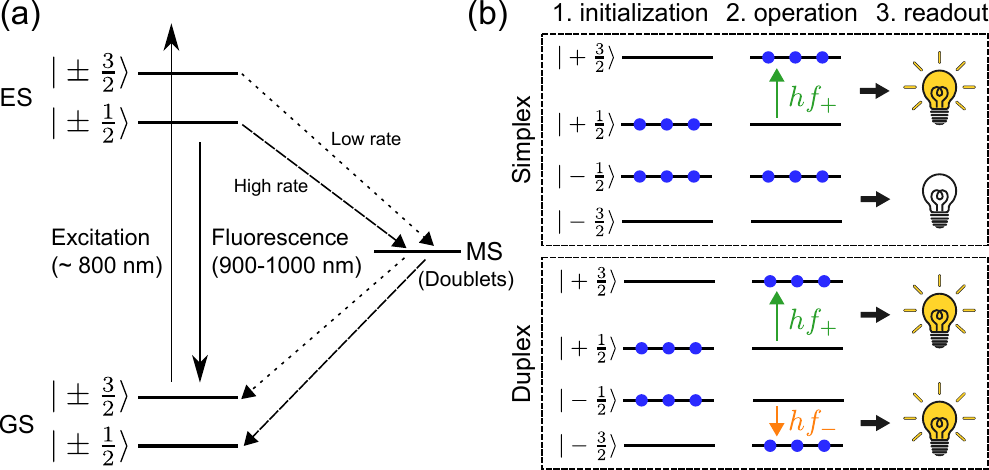}%
  \else
    \includegraphics[width=14cm]{img/Vsi-levels.pdf}%
  \fi
\caption{\label{fig:Vsi-levels} Overview of the proposed method. (a) Energy levels of \Vsi\ in 4H-SiC. (b) Schematic of duplex qubit operation.
  Initialization with non-resonant laser pulse populates $\ket{\pm 1/2}$.
  A qubit operation with MW ($\pi$ pulse) inverts the population between $\ket{\pm 3/2}$ and $\ket{\pm 1/2}$.
  Optical readout is enabled by stronger fluorescence when spin state is in $\ket{\pm 3/2}$.
}
\end{figure}

However, the \Vsi\ qubit has a significant drawback due to its low ODMR contrast at room temperature.
The contrast of a few percent is an order of magnitude lower than that of NV centers in diamond\cite{rondinMagnetometryNitrogenvacancyDefects2014,barrySensitivityOptimizationNVdiamond2020} or divacancy-related centers in SiC\cite{liRoomtemperatureCoherentManipulation2022}, which typically exhibit contrasts of $20-30\%$.
Although the spin-$\frac{3}{2}$ nature of \Vsi\ enables unique sensing capabilities such as vector magnetometry\cite{leeVectorMagnetometryBased2015,niethammerVectorMagnetometryUsing2016} and all-optical metrology using level anti-crossing (LAC)\cite{siminAllOpticalDcNanotesla2016,anisimovOpticalThermometryBased2016,soykalQuantumMetrologySingle2017},
it also imposes a problem in sensing techniques based on pulse ODMR.
Spin state initialization and readout are not as convenient as those of its spin-1 counterpart under non-resonant optical excitation.
That is, the optical spin polarization and ODMR contrast are only available between $\ket{m=\pm 3/2}$ and $\ket{m=\pm 1/2}$ states in the quartet (Fig.\,\ref{fig:Vsi-levels}\,(a)).
After optical excitation from the ground state (GS) to the excited state (ES),
inter-system crossing and non-radiative decay through metastable (MS) doublet states occur at certain rates\cite{widmannCoherentControlSingle2015,dongSpinPolarizationIntersystem2019,liuSiliconVacancycenters2024}.
Optical spin polarization and readout are possible because the rate of this decay path is higher for $\ket{\pm 1/2}$ than $\ket{\pm 3/2}$.
However, a method is yet to be devised for creating the rate difference between $\ket{+ 1/2}$ and $\ket{- 1/2}$ (or $\ket{+ 3/2}$ and $\ket{- 3/2}$).
Thus, the spin is polarized into $\ket{+1/2}$ or $\ket{-1/2}$ with equal probability after optical pumping,
and we cannot optically detect population difference between $\ket{+1/2}$ and $\ket{-1/2}$ (or $\ket{+ 3/2}$ and $\ket{- 3/2}$).
The limited number of sensing demonstrations using \Vsi\ based on pulse ODMR in the literature may be attributed to these issues\cite{castellettoQuantumSystemsSilicon2023}.
A method to overcome these problems is necessary to make \Vsi\ as practical as the NV center in diamond.

To mitigate these problems, we propose a simple technique for simultaneously operating duplex qubits inside the quartet,
that is, $\left\{ \ket{+3/2}, \ket{+1/2} \right\}$ and $\left\{ \ket{-1/2}, \ket{-3/2} \right\}$ as illustrated in Fig.\,\ref{fig:Vsi-levels}\,(b).
If we drive only one qubit of $\left\{ \ket{+3/2}, \ket{+1/2} \right\}$ with a microwave (MW) at resonant frequency $f_+$,
the remaining state $\ket{-1/2}$ is unaffected and becomes background on the optical readout, degrading ODMR contrast.
If both qubits are operated to emit the same signal using two resonant frequencies $f_+$ and $f_-$ ($f_-$ is for $\left\{ \ket{-1/2}, \ket{-3/2} \right\}$),
the total ODMR signal is doubled.
The idea of simultaneously driving multiple electron spin transitions is found in the literature of
diamond NV center magnetometry\cite{zhangDiamondMagnetometryGradiometry2021,herbschlebLowFrequencyQuantumSensing2022} and DC magnetometry based on lock-in-detected ODMR using \Vsi\ in SiC\cite{lekaviciusMagnetometryBasedSiliconVacancy2023},
however, it has never been applied to magnetometry based on qubit control (pulse ODMR) using \Vsi\ in SiC.

In this study, we experimentally demonstrate doubled contrast in pulse ODMR measurements and evaluate its sensitivity in AC magnetometry.
A dense \Vsi\ ensemble in 4H-SiC was used in our experiment.
However, the proposed method can be used for other spin-$\frac{3}{2}$ color centers exhibiting zero-field splitting\cite{krausMagneticFieldTemperature2014}.

\section{Results}

\subsection{Sensing with duplex qubits}

First, we show that duplex qubits can be operated simultaneously in a quantum-sensing scheme.
Let us consider spin Hamiltonian for spin-$\frac{3}{2}$ as follows:
\begin{align}
  H &= D S_z^2 + \hbar \gamma \bm{S} \cdot \bm{B}
\end{align}
where $D, \gamma, \bm{S}$, and $\bm{B}$ are zero-field splitting, electron gyromagnetic ratio (taken positive),
spin operator, and external magnetic field, respectively.
Here, we omitted the terms related to transversal zero-field splitting, external electric field,
and coupling to nuclear spins.
Furthermore, $\bm{B}$ is taken as conventionally found in magnetic resonance literature;
static field $B_0$ is applied along the $z$-axis and MW field $B_1$ is rotating in the $xy$-plane at an angular frequency $\omega$ with initial phase $\phi_1$:
\begin{align}
  \bm{B} = \left( B_1 \cos(\omega t + \phi_1), B_1 \sin(\omega t + \phi_1), B_0 \right)
\end{align}
The rotating frame Hamiltonian $H'$, which cancels the rotation of $\bm{B}_1$, has the following form:
\begin{align}
  H' / \hbar =& \gamma B_1 \cos(\phi_1) S_x + \gamma B_1 \sin(\phi_1) S_y \nonumber \\
              & + \left(D/\hbar\right) S_z^2 + \left(\gamma B_0 - \omega \right) S_z
\end{align}
When upper left (lower right) $2\times 2$ subspace is considered for qubit basis $\left\{ \ket{+3/2}, \ket{+1/2} \right\}$ ($\left\{ \ket{-1/2}, \ket{-3/2} \right\}$), the Hamiltonian can be represented by Pauli matrices $\bm{\sigma}$ as follows:
\begin{align}
  H' / \hbar &\doteq \frac{\bm{\sigma}}{2} \cdot \left(\omega_1 \cos\phi_1,\,\omega_1 \sin\phi_1,\,\Delta \omega\right) \label{eq:Hqubit}
\end{align}
Here $\omega_1 = \sqrt{3} \gamma B_1$, $\Delta \omega = \omega_0 - \omega$,  and
\begin{align}
  \omega_0 =
  \begin{cases}
    \gamma B_0 + 2 D / \hbar & \left(\left\{ \ket{+3/2}, \ket{+1/2} \right\}\right) \\
    \gamma B_0 - 2 D / \hbar & \left(\left\{ \ket{-1/2}, \ket{-3/2} \right\}\right)
  \end{cases}
  \label{eq:omega0}
\end{align}
We disregarded the transition between $\left\{ \ket{+3/2}, \ket{+1/2} \right\}$ and $\left\{ \ket{-1/2}, \ket{-3/2} \right\}$ subspaces
because $T_1$ relaxation is considered negligible and MW fields for such transitions are spectrally isolated;
for example, the angular frequency of transition $\ket{+1/2} \leftrightarrow \ket{-1/2}$ is $\gamma B_0$.
We also assume that $B_0$ is well outside the LAC condition.
We can use these state pairs as qubits because Hamiltonian (\ref{eq:Hqubit}) has the same form as a spin-$\frac{1}{2}$ qubit.
In the qubits, the detuning $\Delta \omega$ results in a phase accumulation (in absence of $\omega_1$, rotation of Bloch vector around $z$ axis),
and an oscillating field $\omega_1 \gg |\Delta \omega|$ drives the Rabi nutation (rotation around an axis perpendicular to $z$).
We call the qubit consisting of $\left\{ \ket{+3/2}, \ket{+1/2} \right\}$ the ``\pqubit", and $\left\{ \ket{-1/2}, \ket{-3/2} \right\}$ the ``\mqubit".

\begin{figure*}
  \includegraphics[width=16cm]{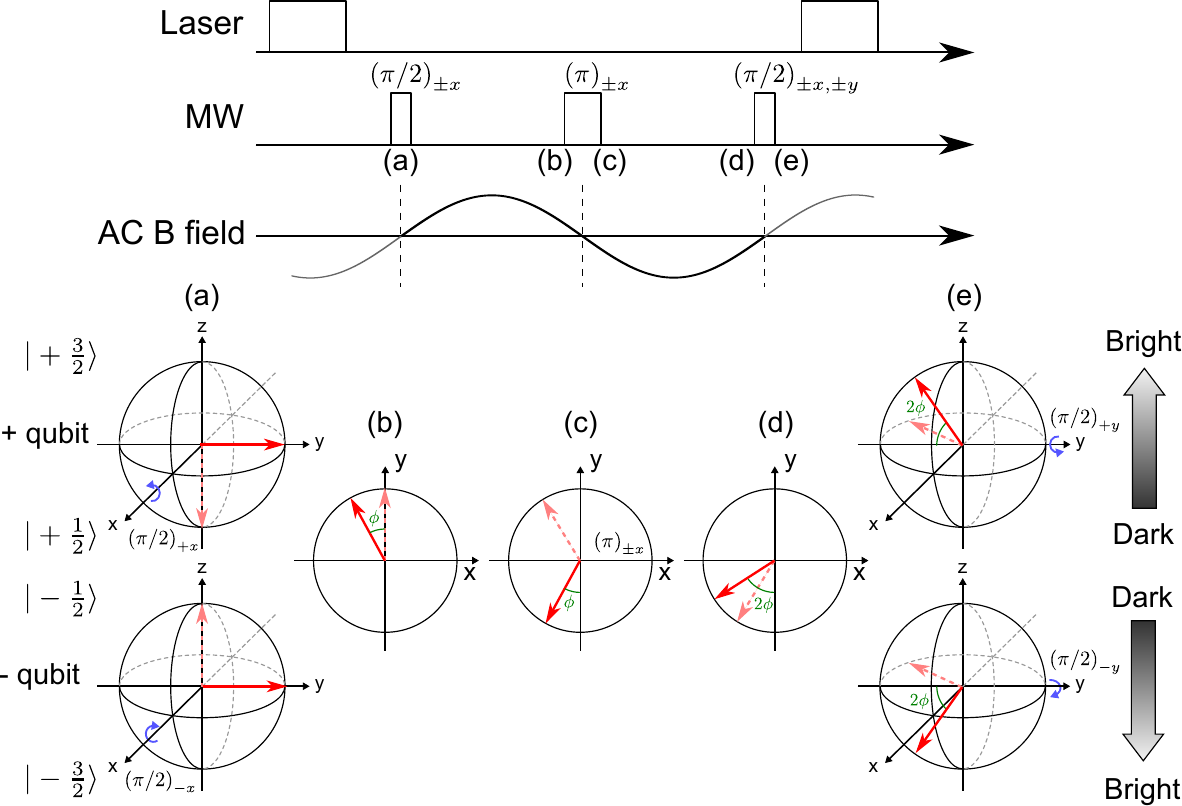}%
\caption{\label{fig:echo-qubits} AC magnetometry based on spin echo using duplex qubits. Upper panel shows the pulse sequence (laser and MW) and AC magnetic field to be measured. Lower panel shows qubit states at each step of the sequence.
(a) Initialization by the laser pulse and the first $\pi/2$ pulse.
(b) Phase accumulation during the first inter-pulse period.
(c) Flipping by the $\pi$ pulse.
(d) Phase accumulation during the second inter-pulse period.
(e) Conversion of phase to population difference by the second $\pi/2$ pulse.
}
\end{figure*}

Let us consider a quantum-sensing method based on pulse ODMR using these two qubits.
Specifically, we use AC magnetometry with a spin-echo sequence as an example,
which is a standard technique for sensing with color center spins.
Fig.\,\ref{fig:echo-qubits} depicts the sensing sequence,
which comprises a laser pulse for qubit initialization, MW pulses for state control,
and another laser pulse for readout.
Two MW pulses with frequencies in resonance with the $+$ and $-$ qubits are
applied simultaneously to control both qubits at the same time.
The MW pulse train is synchronized with the AC magnetic field $\tilde{B}_z(t)$ to be measured,
the sign of which is inverted between the two inter-pulse periods (see top of Fig.\,\ref{fig:echo-qubits}).
With unknown magnetic field $\tilde{B}_z(t)$ introduced to the systems
with exact resonance condition $\omega = \gamma B_0 \pm 2 D / \hbar$,
the detuning becomes $\Delta \omega (t) = \gamma \tilde{B}_z(t)$ for both qubits.
The AC magnetic field is measured through the qubit phase accumulation caused by this detuning.

The first laser pulse initializes the spin states into $\ket{\pm 1/2}$,
and subsequent $\pi / 2$ pulse generates superposition states in qubits, $\left(\ket{+3/2} + i \ket{+1/2} \right) / \sqrt{2}$ or $\left(\ket{-1/2} + i \ket{-3/2} \right) / \sqrt{2}$ (Fig.\,\ref{fig:echo-qubits}\,(a)).
Given MW phase $\phi_1$, we set the coordinate of the Bloch sphere so that the $\pi / 2$ pulse rotates the Bloch vector around $+x$ axis ($-x$ axis) for \pqubit (\mqubit).
The Bloch vectors of both qubits are then initialized as pointing in the $+y$ direction.
Figure \ref{fig:echo-qubits} (b)-(d) illustrate the process of phase accumulation
during the spin echo sequence: a free evolution in first inter-pulse period,
state flipping by a $\pi$ pulse, and another free evolution in the second period, respectively.
By applying the $\pi$ pulse, the effect of which is identical for both $+$ and $-$ qubits,
the dynamics caused by $\tilde{B}_z(t)$ results in a measurable phase
($\phi$ in an inter-pulse period, $2\phi$ in total)
because of sign alteration between first and second inter-pulse periods.
The effect of time-independent perturbation is cancelled by this sequence.
The second $\pi / 2$ pulse is used to convert the accumulated phase $2\phi$ into a population difference
($z$ coordinate in Bloch space) to enable optical readout.
Here, the MW phase $\phi_1$ is adjusted so that the Bloch vector rotates around the $\pm x$ or $\pm y$ axes.
Figure \ref{fig:echo-qubits}\,(e) depicts the motions of the Bloch vectors by $\left(\pi / 2\right)_{\pm y}$ pulses for \pmqubits,
which results in Bloch vectors with $z$ component $\pm \sin(2\phi)$.
Note that the $+$ qubit gives stronger fluorescence when it points in the $+$ direction, and vice versa.
Therefore, we expect the same signal (change in fluorescence intensity caused by $\tilde{B}_z(t)$)
from both qubits, resulting in a stronger signal contrast.
If we assume $\left(\pi / 2\right)_{\pm x}$ pulse instead of $\left(\pi / 2\right)_{\pm y}$ pulse,
the $z$ components of final Bloch vectors become $\mp \cos(2\phi)$,
which could also be used for enhancing the contrast.

So far, we focused on measuring magnetic field which couples to first order in $\bm{S}$ in the Hamiltonian.
If we wish to measure quantities coupling to even-order in ${\bm S}$, such as electric field,
we need to make a slight adjustment in the phase of the second $\pi / 2$ pulse.
For instance, electric field along the $z$ axis gives rise to a frequency shift with opposite signs
in the two qubits.
A shift in zero-field splitting $D$, which can be caused by varying temperature or lattice distortion,
also results in a perturbation of the opposite sign.
The accumulated phase $\phi$ by such perturbations is also opposite.
This phenomenon does not affect the $\left(\pi / 2\right)_{\pm x}$ pulse readout
because cosine is an even function.
By contrast, it affects the sine-shape signals from the $\left(\pi / 2\right)_{\pm y}$ pulse readout;
the $z$ components of final Bloch vectors become identical (e.g., $\sin(2\phi)$) for both $+$ and $-$ qubits and
changes in fluorescence intensities cancel each other out.
However, we can avoid the cancellation by adjusting the $\pi / 2$ pulse phase,
for example, by using $\left(\pi / 2\right)_{+y}$ pulse for both qubits.

\subsection{CW-ODMR}

\begin{figure}
  \includegraphics[width=8cm]{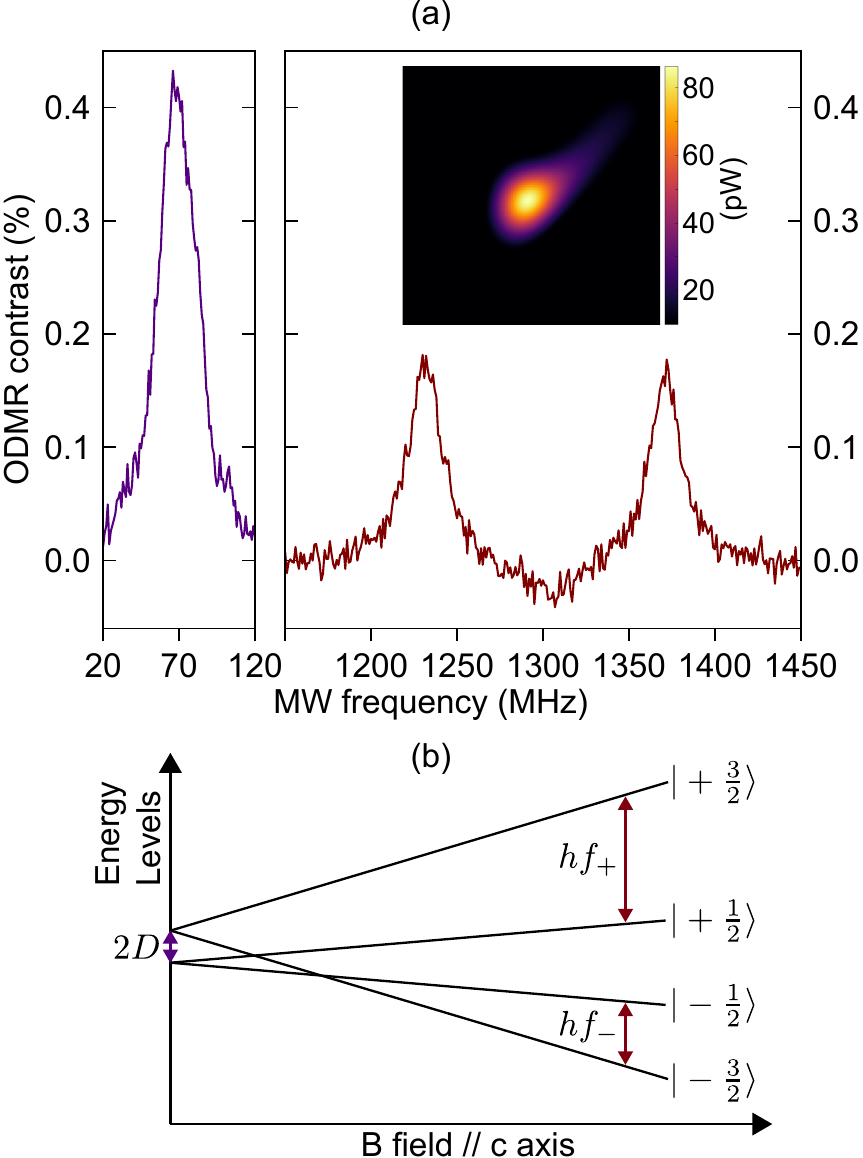}%
\caption{\label{fig:cfm-cw} CW-ODMR characterization.
  (a) CW-ODMR spectra at zero / 46 mT magnetic field along the $c$-axis.
  (inset) Confocal micrograph of the \Vsi\ ensemble. Size: $6\times6$ \micronsq.
  (b) Schematic of the \Vsi\ ground state spin energies under magnetic field along the $c$-axis. }
\end{figure}

The experiments were performed using a home-built confocal microscope and
a dense \Vsi\ cluster (Fig.\,\ref{fig:cfm-cw} inset, estimated density $10^{18}\ \mathrm{cm}^{-3}$, volume 0.5 $\mathrm{\mu m}^3$) created in n-type 4H-SiC (see Methods for details).

Fig.\,\ref{fig:cfm-cw}\,(a) depicts the Continuous Wave (CW) ODMR spectra of the \Vsi\ ensemble.
The spectra were obtained by recording the fluorescence intensity with ($I_1$) and without ($I_0$) MW irradiation, while performing MW frequency sweeping.
Presented ODMR contrast is defined as $(I_1  - I_0) / I_0$.
At zero field, the ground state spin levels $\ket{\pm3/2}$ and $\ket{\pm1/2}$ are both degenerate, and
we observe a single ODMR line at $2D/h \simeq 70 \mathrm{MHz}$ corresponding to magnetic resonance between these states.
The degeneracy is lifted by applying magnetic field $B_0$ along
the $z$ axis (crystallographic $c$ axis);
two main ODMR lines are observed at $f_{\pm} = \gamma B_0/(2 \pi) \pm 2 D/h$ (Fig.\,\ref{fig:cfm-cw}\,(b)).
These frequencies correspond to \pmqubits\ (transitions $\ket{+3/2} \leftrightarrow \ket{+1/2}$ and $\ket{-1/2} \leftrightarrow \ket{-3/2}$) as presented in Eq.\,(\ref{eq:omega0}).
Both spectra were recorded with the same MW power (forward power measured between the amplifier and the sample) and
exhibited contrasts well below saturation, which was approximately 45\% of the saturated value.
However, the contrast between the two spectra differed considerably;
the peak height of the spectrum with the magnetic field was approximately half of that without the magnetic field.
This result is consistent with previous studies\cite{carterSpinCoherenceEcho2015,abrahamNanoteslaMagnetometrySilicon2021}, and result of driving single magnetic resonance (e.g., $\ket{+3/2} \leftrightarrow \ket{+1/2}$ at $f_+$)
while optical pumping is polarizing the state into $\ket{\pm 1/ 2}$ (population in $\ket{-1/2}$ is a background at $f_+$).

\subsection{Contrast Enhancement in Pulse ODMR}

\begin{figure*}
  \includegraphics[width=16cm]{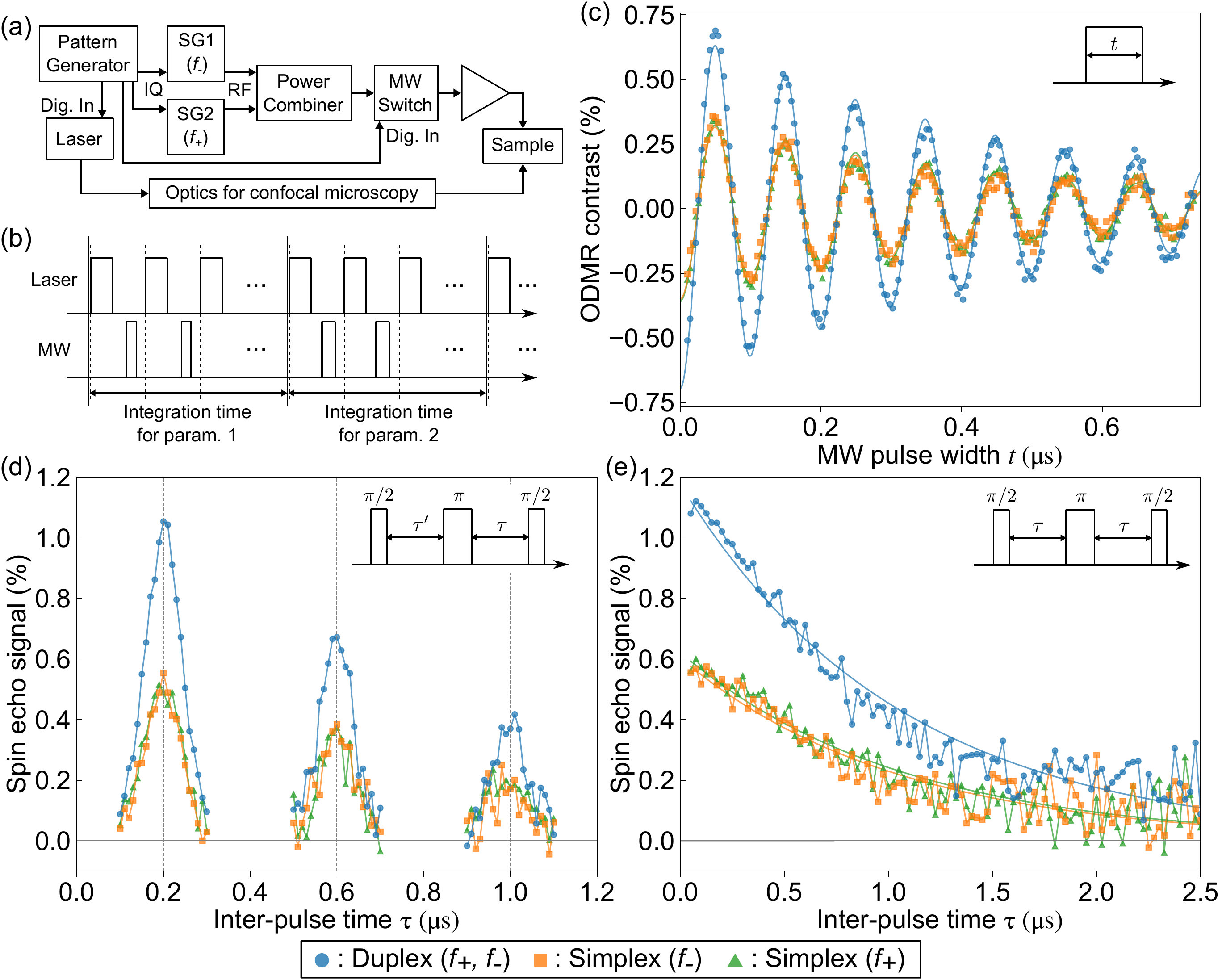}%
\caption{\label{fig:rabi-echo} Pulse ODMR characterization. (a) Schematic of the setup for pulse ODMR.
  (b) Repeated pulse sequence. (c) Rabi oscillation. (d) Spin echo signals at $\tau' = 0.2, 0.6, 1.0\ \mathrm{\mu s}$.
  (e) Spin echo envelope. }
\end{figure*}

After identifying the basic characteristics, we performed pulse ODMR characterizations using duplex qubits.
Figure \ref{fig:rabi-echo}\,(a) depicts a schematic of the experimental setup for the pulse ODMR.
The timing signal was generated using a pulse-pattern generator (Tektronix DTG5274).
A laser pulse was generated using the digital modulation function of the diode laser.
Two signal generators (SGs, Anritsu MG3710E and Keysight N5172B) were used to synthesize MW frequencies $f_{\pm}$.
The phase of MW signal was modulated using the IQ modulation function of the SGs.
The generated signals were combined using a power combiner (Mini-Circuits ZN2PD-4R753+)
before being input into the switch, amplifier, and the sample, as described in the Methods.

The pulse sequence comprised a laser pulse for spin (qubit) state initialization, a MW pulse pattern for operation,
and another laser pulse for state readout.
We set the laser pulse width and delay time before MW pulses as 0.5 and 0.7 \microsec, respectively.
The photo-receiver utilized had a narrow bandwidth (1.1 kHz); hence, we could not acquire the time-resolved fluorescence for each laser pulse.
Instead, we recorded the average fluorescence intensity during repeated sequences (Fig.\,\ref{fig:rabi-echo}\,(b)).
The integration time for each acquisition was fixed at 1/60 s to cancel power-line noise (60 Hz).

First, the MW power was calibrated for duplex qubit operation by observing the Rabi oscillation.
The Rabi frequencies $\omega_1$ should be matched for \pmqubits\ to operate them simultaneously.
However, $\omega_1$ (MW amplitude $B_1$) for two frequencies can diverge even when the same power is output from SG1 and SG2,
because of the frequency characteristics of the components (combiner, amplifier, cable, etc.).
Therefore, we tuned the SG output power so that each Rabi frequency with simplex qubit operations (only one of $f_{+}$ or $f_{-}$ is output)
reaches the target value of 10 MHz.
Subsequently, both SGs were turned on to confirm the Rabi oscillation with the duplex qubit operation.
Figure \ref{fig:rabi-echo}\,(c) depicts the Rabi oscillations after calibration.
The two characteristics of simplex operations coincide, and the duplex operation resulted in a larger signal amplitude.
Here, the ODMR contrast was defined to be $C(t) = \left(I(t)-\bar{I}\right) / \bar{I}$, where $\bar{I}$ is the mean of intensity $I(t)$.
The signals were fitted using the following model:
\begin{equation}
  C(t) = - A_R \cos \left(\omega_1 t\right) e^{- t / T_2^{\star}} \label{eq:rabi-fit}
\end{equation}
The peak-to-peak amplitudes were $2 A_R = 0.71 \%$ for the simplex operation (mean of two characteristics)
and $2 A_R = 1.40 \%$ for the duplex operation, revealing a signal gain of 1.97, which is close to the ideal value of 2.

Next, the spin-echo measurements were performed to characterize the coherence properties of the qubits.
The sequence comprised standard $\pi/2-\pi-\pi/2$ pulses (widths of $25-50-25$ ns at the calibrated power).
The spin-echo signals were recorded in a complementary manner;
one data ($I_a$) was collected with a second $\pi/2$ pulse phase (e.g., $\left(\pi / 2\right)_{+x}$)
as well as another data ($I_b$) with flipped phase ($\left(\pi / 2\right)_{-x}$).
Then, the spin-echo signal was defined as $F = \left(I_a - I_b\right) / \bar{I} = 2 \left(I_a - I_b\right) / \left(I_a + I_b\right)$.

Fig.\,\ref{fig:rabi-echo}\,(d) depicts the observed spin-echo signals.
Here, first inter-pulse time $\tau'$ was fixed, and second time $\tau$ was swept as a parameter.
Peak-shape signal centered at $\tau'=\tau$ is clearly observed and duplex measurement results in enhanced signal as in Rabi oscillation.
The envelope of echo peak is depicted in Fig.\,\ref{fig:rabi-echo}\,(e), where two inter-pulse times were identical.
These signals are fit with simple exponential decay model $F(\tau) \propto e^{- 2 \tau / T_2}$,
and a coherence time of $T_2 = 2.1\,\mathrm{\mu s}$ is extracted.
This coherence time is not satisfactory compared with the tens or hundreds of microseconds reported for low-density \Vsi \cite{widmannCoherentControlSingle2015,carterSpinCoherenceEcho2015,kasperInfluenceIrradiationDefect2020}, however,
this parameter is a reasonable value considering that estimated \Vsi\ density in our sample is high ($10^{18}\ \mathrm{cm}^{-3}$) \cite{kasperInfluenceIrradiationDefect2020}.

\subsection{Sensitivity Enhancement in AC magnetometry}

\begin{figure*}
  \includegraphics[width=16cm]{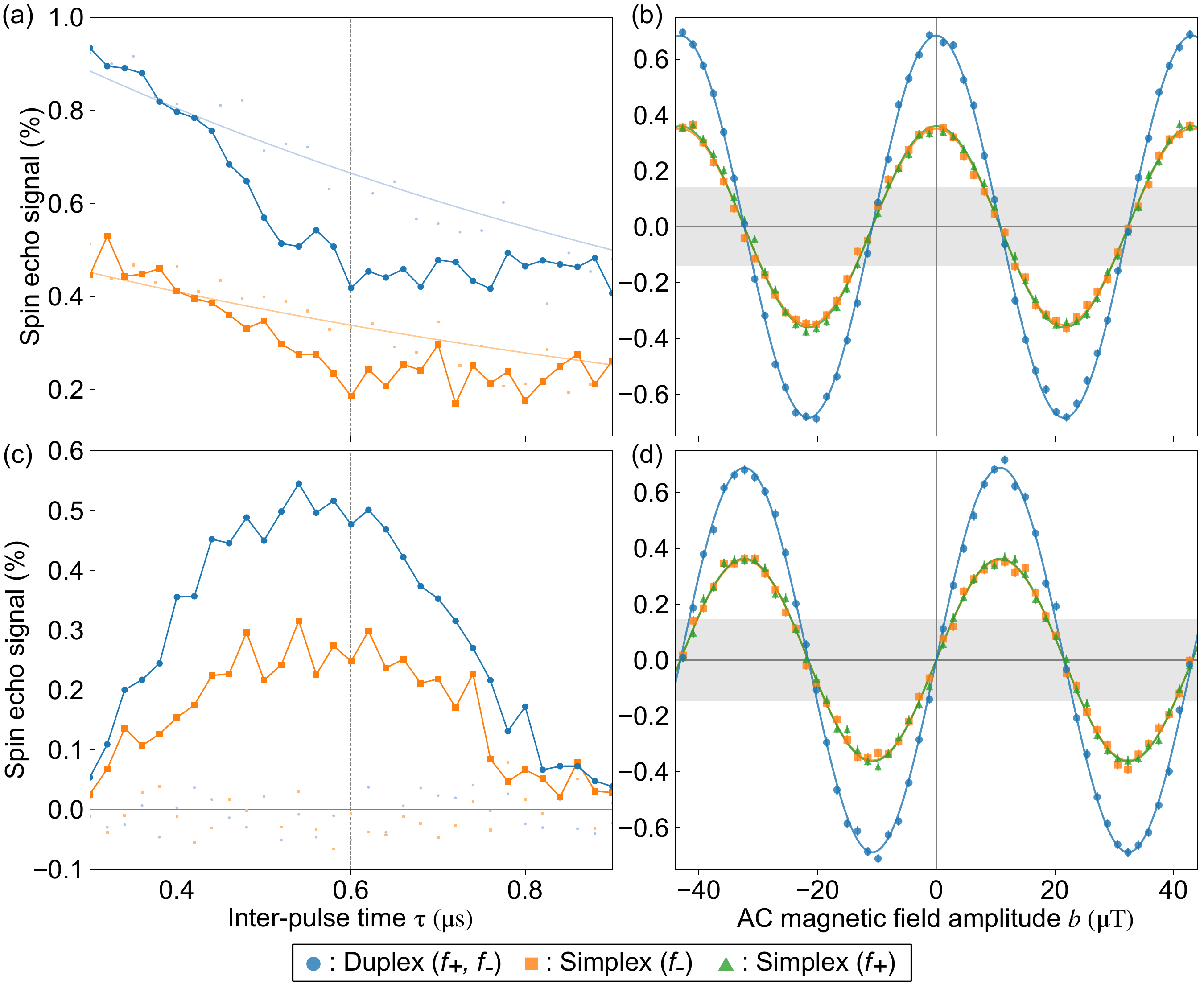}%
\caption{\label{fig:mag} Spin echo AC magnetometry.
  (a) $\tau$-dependence (frequency characteristics) at $b=5.75\ \mathrm{\mu T}$ with $\left(\pi / 2\right)_{\pm x}$ readout.
  (b) $b$-dependence (response characteristics) at $\tau=0.6\ \mathrm{\mu s}$ with $\left(\pi / 2\right)_{\pm x}$ readout.
  (c) $\tau$-dependence with $\left(\pi / 2\right)_{\pm y}$ readout.
  (d) $b$-dependence with $\left(\pi / 2\right)_{\pm y}$ readout.
  Small dots in (a) and (c) depict the baseline without AC test signal ($b=0$).
  One of simplex data ($f_{+}$) is omitted in (a) and (c) to ensure the readability.
  Grey bands in (b) and (d) represent noise levels in 1 s accumulation ($\pm \sigma_F(1)$ in Table\,\ref{tab:mag-result}).
  Error bars indicate standard deviations for each data point (80 s accumulation).
}
\end{figure*}

Finally, we demonstrate AC magnetometry based on a spin-echo sequence.
The test AC signal was generated using a function generator and applied to the \Vsi\ ensemble using a 16-turn coil placed beneath the SiC substrate.
The sinusoidal signal was synchronized with the spin-echo pulse sequence (Fig.\,\ref{fig:echo-qubits}).
The frequency of the test signal, 769.23 kHz, was chosen to perfectly synchronize to the pulse sequence at $\tau = 0.6\ \mathrm{\mu s}$.
Here, the finite widths of the MW pulses are considered.
First, the $\tau$ dependence of spin echo modulation (frequency characteristics) is confirmed.
Fig.\,\ref{fig:mag}\,(a) and (c) depict these characteristics
with the second $\pi / 2$ pulse (readout) phase of $\pm x$ and $\pm y$, respectively.
For the $\pm x$ readout, the response is a dip centered at $\tau = 0.6\ \mathrm{\mu s}$
from the baseline of unmodulated envelope (Fig.\,\ref{fig:rabi-echo}\,(e)).
By contrast, the baseline is $F=0$ for $\pm y$ readout because sine-shape response is expected instead of cosine-shape.
A little asymmetry of the peak shape was observed due to decoherence, that is, short $\tau$ results in enhanced signal scale compared with longer $\tau$.
Critically, the duplex measurement revealed a signal amplitude
which was two times that of the simplex measurement.

\begin{figure}
  \if\twocol1
    \includegraphics[width=8.5cm]{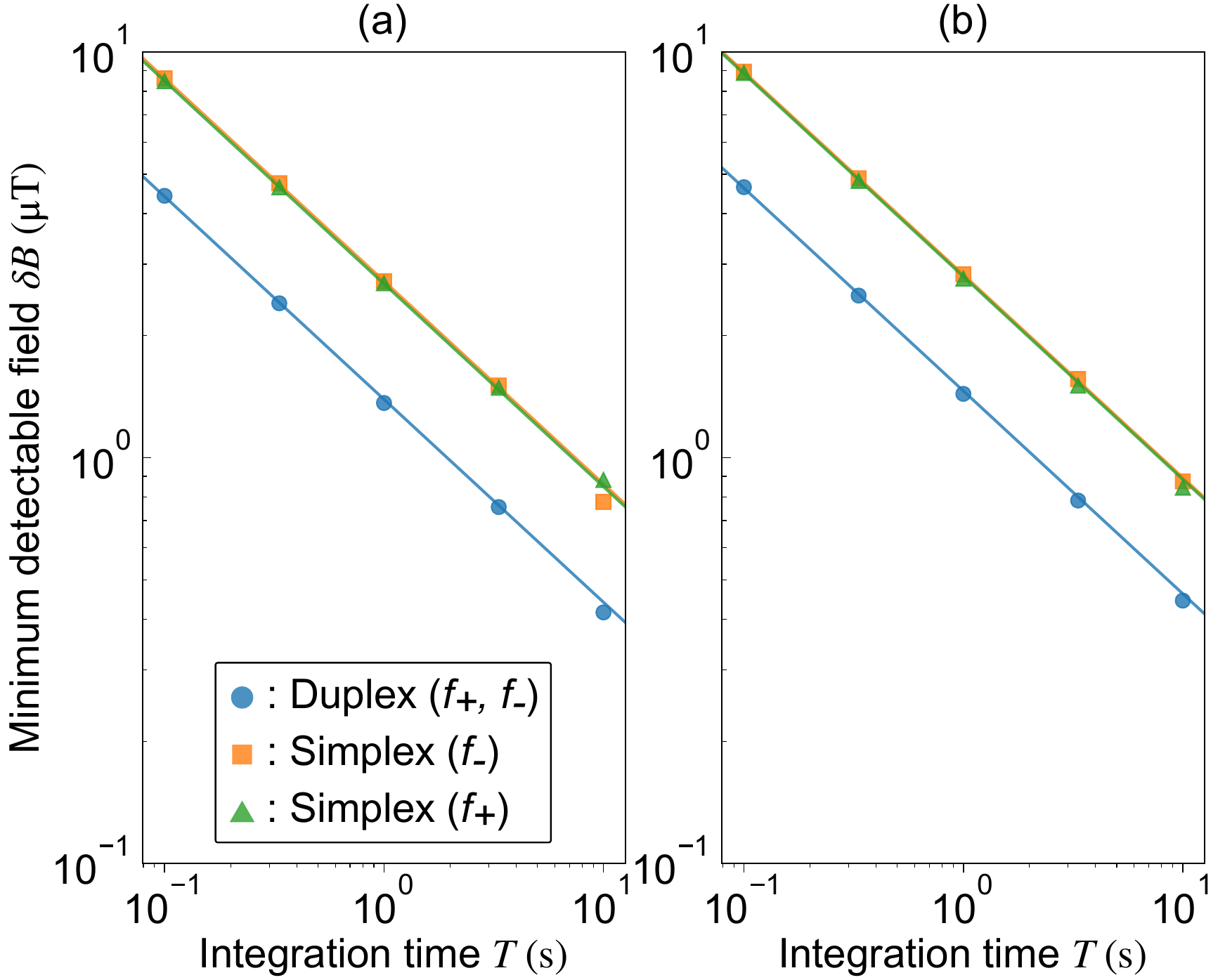}%
  \else
    \includegraphics[width=12.5cm]{img/sens_scaling_ed.pdf}%
  \fi
\caption{\label{fig:sens-scaling} Scaling of AC magnetometry sensitivity.
  (a) $\left(\pi / 2\right)_{\pm x}$ readout. (b) $\left(\pi / 2\right)_{\pm y}$ readout.
  Lines are fitting results to the expression $\delta B(T) = \eta / \sqrt{T}$. }
\end{figure}

Fig.\,\ref{fig:mag} (b) and (d) depict the dependences on the AC field amplitude at $\tau=0.6\ \mathrm{\mu s}$.
When the AC field is expressed as $\tilde{B}_z(t) = b \sin (2 \pi \nu t)$, the accumulated phase $\phi$ in an inter-pulse period becomes the following:
\begin{align}
  \phi = \int_0^{1/2\nu} \gamma \tilde{B}_z(t) dt = \frac{\gamma b}{\pi \nu} \label{eq:phi}
\end{align}
Therefore, the spin echo signals are expressed as follows:
\begin{align}
  F_x(b) = A \cos \left(\frac{2\gamma b}{\pi \nu}\right),\ F_y(b) = A \sin \left(\frac{2\gamma b}{\pi \nu}\right)
\end{align}
where $F_x$ and $F_y$ denote signals by the $\pm x$ and $\pm y$ readout.
The data fit these expressions well (Fig.\,\ref{fig:mag} (b) and (d)).
Sensitivity is defined as the minimum detectable field within the unit acquisition time,
which can be expressed as follows\cite{phamEnhancedMetrologyUsing2012,naydenovDynamicalDecouplingSingleelectron2011,delangeSingleSpinMagnetometryMultipulse2011}:
\begin{align}
  \eta &= \frac{\sigma_F(1)}{\delta F} &\left(\mathrm{T/\sqrt{Hz}}\right)\\
  \delta F &= \max \left(\frac{dF}{db}\right) = \frac{2 A \gamma}{\pi \nu} &\left(\mathrm{T^{-1}}\right) \label{eq:sensitivity}
\end{align}
Here, the maximum slope of response, $\delta F$, is evaluated at $b = \pi^2 \nu / (4 \gamma b)$ ($\pm x$ readout) or $b = 0$ ($\pm y$ readout).
The noise level $\sigma_F(1)$ is a fluctuation of $F$ in 1 s acquisition time.
To quantify $\eta$ experimentally, the slope is computed using fitting parameter $A$ and
$\sigma_F$ is estimated by sampling standard deviation from repeated measurements.
The dependence of $\sigma_F$ on integration time $T$ is confirmed to be $\sigma_F(T) = \sigma_F(1)/\sqrt{T}$,
whereas the main noise source in our experiment is yet to be identified
(candidates include laser power fluctuation, amplifier in the photo-receiver, or A/D converter).
Fig.\,\ref{fig:sens-scaling} shows the $T$-dependence of minimum detectable field, $\delta B(T) = \sigma_F(T) / \delta F = \eta / \sqrt{T}$.

Table\,\ref{tab:mag-result} summarizes the results of the sensitivity evaluation.
The most significant difference was between duplex and simplex operations;
the doubled ODMR contrast in duplex operations resulted in doubled $A$ and $\delta F$.
The $\sigma_F$ marked similar values for all experiments,
which indicates that the main noise source is in the light source or detection system.
Therefore, the contrast improvement is directly reflected in sensitivity $\eta$ as well.
The gains in sensitivity are 1.99 ($\pm x$ readout) and 1.94 ($\pm y$ readout),
which are close to ideal value 2.

\begin{table}[tb]
  \caption{\label{tab:mag-result} Summary of sensitivity evaluation in AC magnetometry.
  Phase indicates the phase of the second $\pi / 2$ pulse.}
  \begin{ruledtabular}
  \begin{tabular}{cl|ccc}
  Phase & Operation & $\delta F$     & $\sigma_F(1)$ & $\eta$ \\
        &           & $\mathrm{(\%/\mu T)}$ & $\mathrm{(\%/\sqrt{Hz})}$ & $\mathrm{(\mu T/\sqrt{Hz})}$ \\ \hline
  $\pm x$   & Simplex ($f_{-}$)       & 0.0513  & 0.140 & 2.73 \\
  $\pm x$   & Simplex ($f_{+}$)       & 0.0525  & 0.141 & 2.69 \\
  $\pm x$   & Duplex ($f_{+}, f_{-}$) & 0.0998  & 0.136 & 1.36 \\
  $\pm y$   & Simplex ($f_{-}$)       & 0.0524  & 0.148 & 2.83 \\
  $\pm y$   & Simplex ($f_{+}$)       & 0.0529  & 0.146 & 2.76 \\
  $\pm y$   & Duplex ($f_{+}, f_{-}$) & 0.1004  & 0.144 & 1.44 \\
  \end{tabular}
  \end{ruledtabular}
\end{table}

\section{Discussion}

In this study, we proposed a method to operate duplex qubits in spin-$\frac{3}{2}$ quartet simultaneously
by applying MW pulses containing two resonant frequencies.
The ODMR contrast and sensitivity were doubled by this method compared with that of the conventional simplex operation.
Although we used \Vsi\ in 4H-SiC for our experiments, the technique can be extended to other color-center qubits
with spin-$\frac{3}{2}$ (or greater $S$) exhibiting zero-field splitting\cite{krausMagneticFieldTemperature2014}.

We demonstrated an AC magnetometry sensitivity $1.4\ \mathrm{\mu T / \sqrt{Hz}}$ with a dense \Vsi\ cluster within the 0.5 $\mathrm{\mu m}^3$ volume.
However, the utility of the proposed method is independent of the system size;
the method can be applied to achieve improved sensitivity with ensembles in larger volumes\cite{lekaviciusMagnetometryBasedSiliconVacancy2023}
as well as nanoscale sensing with a single defect.
The method can also be applied to DC magnetometry (Ramsey interferometry)
or to the sensing of quantities other than magnetic fields.
It is noteworthy that Lekavicius \textit{et al}. has proposed another method, the use of $\left\{ \ket{+1/2}, \ket{-1/2} \right\}$ qubit basis, to extract better $T_2^{\star}$ for DC magnetometry\cite{lekaviciusMagnetometryBasedSiliconVacancy2023,lekaviciusOrdersMagnitudeImprovement2022}.
There is a trade-off between $T_2^{\star}$ and contrast in the two methods.
Since enhancement of $T_2^{\star}$ could depend on the material, defect density, or sensor volume,
experimental investigations will be required to determine which method results
in a better DC sensitivity for a given system.

Fundamental performances can be enhanced in the future using several approaches.
The \Vsi\ density could be optimized to achieve a superior balance between the fluorescence intensity and coherence time.
Material engineering techniques, such as post-irradiation annealing\cite{kasperInfluenceIrradiationDefect2020} or isotope concentration\cite{lekaviciusOrdersMagnitudeImprovement2022,lekaviciusMagnetometryBasedSiliconVacancy2023}, can eliminate unwanted paramagnetic defects or nuclear spins
to increase the coherence time.
Dynamical decoupling sequences can be used to extract superior coherence and frequency selectivity\cite{naydenovDynamicalDecouplingSingleelectron2011,delangeSingleSpinMagnetometryMultipulse2011}.

\section{Methods}
\subsection{Confocal microscopy and ODMR measurements}

All experiments were conducted using a home-built confocal microscope at room temperature.
A 785-nm excitation laser light (Omicron LuxX+ 785-200) was tuned, reflected by a dichroic mirror (850 nm cut-on),
and focused onto the sample using an objective lens (Olympus LCPLN50XIR) mounted on a three-axis piezo stage.
The fluorescence was collected by the same objective,
passed through the dichroic mirror, a notch filter (785 nm), a long-pass filter (900 nm), and a pinhole (20\micron).
A photo-receiver (FEMTO OE-200-IN1, gain set at $10^{11}$ V/W) was used to convert the fluorescence intensity into an analog voltage signal,
which was recorded using a data acquisition device (National Instruments USB-6363).
The excitation power was 5 mW for all the experiments.

The MW for ODMR was synthesized by SGs,
passed through a MW switch (Mini-Circuits ZYSWA-2-50DR+), amplified (ZHL-20W-202-S+),
inputted through a thin Cu wire (20\micron\ diameter) on the substrate, and subsequently terminated.
A static magnetic field was applied using a permanent (SmCo) magnet placed near the sample.
The instruments were controlled using MAHOS software\cite{taharaMAHOSMeasurementAutomation2023}.

\subsection{Sample preparation}
\Vsi\ was prepared by implanting He$^{+}$ ions into a 4H-SiC substrate
that has a 6.1\micron-thick n-type epi-layer with a nitrogen impurity density of $10^{16}\ \mathrm{cm}^{-3}$.
An acceleration energy of 0.5 MeV resulted in a stopping range distribution of 1\micron\ centered at 1\micron\ depth
from the surface (estimated by SRIM simulation).
The ions were focused onto a spot of diameter 1\micron\ \cite{krausThreeDimensionalProtonBeam2017},
that is confirmed in the confocal image in Fig.\,\ref{fig:cfm-cw}.
Thus, the shape of the \Vsi\ cluster was approximately a sphere of diameter 1\micron\ and its volume was 0.5 $\mathrm{\mu m}^3$.
Each spot was irradiated with $6\times10^{5}$ ions.
Considering the \Vsi\ creation yield of 0.1 per proton (H$^{+}$) reported in \cite{krausThreeDimensionalProtonBeam2017}, and
difference in the vacancy creation rate between He$^{+}$ and H$^{+}$ by a factor of 8 (SRIM simulation),
the number of \Vsi\ inside this cluster was estimated to be $4.8\times10^{5}$.
Equivalently, the volume density of \Vsi\ was estimated to be $1\times10^{18}\ \mathrm{cm}^{-3}$.
Post-irradiation annealing was not performed for this sample.


\section{Acknowledgments}
\begin{acknowledgments}

This work was performed for a research theme
``Environment development for practical use of solid-state quantum sensors: towards social implementation"
in Council for Science, Technology and Innovation (CSTI),
Cross-ministerial Strategic Innovation Promotion Program (SIP),
``Promoting application of advanced quantum technologies to social challenges" (Funding agency: QST).

We thank Hidekazu Tsuchida for the epitaxial growth of SiC.

\end{acknowledgments}







\bibliography{}

\begin{thebibliography}{37}%
\makeatletter
\providecommand \@ifxundefined [1]{%
 \@ifx{#1\undefined}
}%
\providecommand \@ifnum [1]{%
 \ifnum #1\expandafter \@firstoftwo
 \else \expandafter \@secondoftwo
 \fi
}%
\providecommand \@ifx [1]{%
 \ifx #1\expandafter \@firstoftwo
 \else \expandafter \@secondoftwo
 \fi
}%
\providecommand \natexlab [1]{#1}%
\providecommand \enquote  [1]{``#1''}%
\providecommand \bibnamefont  [1]{#1}%
\providecommand \bibfnamefont [1]{#1}%
\providecommand \citenamefont [1]{#1}%
\providecommand \href@noop [0]{\@secondoftwo}%
\providecommand \href [0]{\begingroup \@sanitize@url \@href}%
\providecommand \@href[1]{\@@startlink{#1}\@@href}%
\providecommand \@@href[1]{\endgroup#1\@@endlink}%
\providecommand \@sanitize@url [0]{\catcode `\\12\catcode `\$12\catcode
  `\&12\catcode `\#12\catcode `\^12\catcode `\_12\catcode `\%12\relax}%
\providecommand \@@startlink[1]{}%
\providecommand \@@endlink[0]{}%
\providecommand \url  [0]{\begingroup\@sanitize@url \@url }%
\providecommand \@url [1]{\endgroup\@href {#1}{\urlprefix }}%
\providecommand \urlprefix  [0]{URL }%
\providecommand \Eprint [0]{\href }%
\providecommand \doibase [0]{https://doi.org/}%
\providecommand \selectlanguage [0]{\@gobble}%
\providecommand \bibinfo  [0]{\@secondoftwo}%
\providecommand \bibfield  [0]{\@secondoftwo}%
\providecommand \translation [1]{[#1]}%
\providecommand \BibitemOpen [0]{}%
\providecommand \bibitemStop [0]{}%
\providecommand \bibitemNoStop [0]{.\EOS\space}%
\providecommand \EOS [0]{\spacefactor3000\relax}%
\providecommand \BibitemShut  [1]{\csname bibitem#1\endcsname}%
\let\auto@bib@innerbib\@empty
\bibitem [{\citenamefont {Balasubramanian}\ \emph {et~al.}(2008)\citenamefont
  {Balasubramanian}, \citenamefont {Chan}, \citenamefont {Kolesov},
  \citenamefont {{Al-Hmoud}}, \citenamefont {Tisler}, \citenamefont {Shin},
  \citenamefont {Kim}, \citenamefont {Wojcik}, \citenamefont {Hemmer},
  \citenamefont {Krueger}, \citenamefont {Hanke}, \citenamefont
  {Leitenstorfer}, \citenamefont {Bratschitsch}, \citenamefont {Jelezko},\ and\
  \citenamefont {Wrachtrup}}]{balasubramanianNanoscaleImagingMagnetometry2008}%
  \BibitemOpen
  \bibfield  {author} {\bibinfo {author} {\bibfnamefont {G.}~\bibnamefont
  {Balasubramanian}}, \bibinfo {author} {\bibfnamefont {I.~Y.}\ \bibnamefont
  {Chan}}, \bibinfo {author} {\bibfnamefont {R.}~\bibnamefont {Kolesov}},
  \bibinfo {author} {\bibfnamefont {M.}~\bibnamefont {{Al-Hmoud}}}, \bibinfo
  {author} {\bibfnamefont {J.}~\bibnamefont {Tisler}}, \bibinfo {author}
  {\bibfnamefont {C.}~\bibnamefont {Shin}}, \bibinfo {author} {\bibfnamefont
  {C.}~\bibnamefont {Kim}}, \bibinfo {author} {\bibfnamefont {A.}~\bibnamefont
  {Wojcik}}, \bibinfo {author} {\bibfnamefont {P.~R.}\ \bibnamefont {Hemmer}},
  \bibinfo {author} {\bibfnamefont {A.}~\bibnamefont {Krueger}}, \bibinfo
  {author} {\bibfnamefont {T.}~\bibnamefont {Hanke}}, \bibinfo {author}
  {\bibfnamefont {A.}~\bibnamefont {Leitenstorfer}}, \bibinfo {author}
  {\bibfnamefont {R.}~\bibnamefont {Bratschitsch}}, \bibinfo {author}
  {\bibfnamefont {F.}~\bibnamefont {Jelezko}},\ and\ \bibinfo {author}
  {\bibfnamefont {J.}~\bibnamefont {Wrachtrup}},\ }\bibfield  {title} {\bibinfo
  {title} {Nanoscale imaging magnetometry with diamond spins under ambient
  conditions},\ }\href {https://doi.org/10.1038/nature07278} {\bibfield
  {journal} {\bibinfo  {journal} {Nature}\ }\textbf {\bibinfo {volume} {455}},\
  \bibinfo {pages} {648} (\bibinfo {year} {2008})}\BibitemShut {NoStop}%
\bibitem [{\citenamefont {Maze}\ \emph {et~al.}(2008)\citenamefont {Maze},
  \citenamefont {Stanwix}, \citenamefont {Hodges}, \citenamefont {Hong},
  \citenamefont {Taylor}, \citenamefont {Cappellaro}, \citenamefont {Jiang},
  \citenamefont {Dutt}, \citenamefont {Togan}, \citenamefont {Zibrov},
  \citenamefont {Yacoby}, \citenamefont {Walsworth},\ and\ \citenamefont
  {Lukin}}]{mazeNanoscaleMagneticSensing2008}%
  \BibitemOpen
  \bibfield  {author} {\bibinfo {author} {\bibfnamefont {J.~R.}\ \bibnamefont
  {Maze}}, \bibinfo {author} {\bibfnamefont {P.~L.}\ \bibnamefont {Stanwix}},
  \bibinfo {author} {\bibfnamefont {J.~S.}\ \bibnamefont {Hodges}}, \bibinfo
  {author} {\bibfnamefont {S.}~\bibnamefont {Hong}}, \bibinfo {author}
  {\bibfnamefont {J.~M.}\ \bibnamefont {Taylor}}, \bibinfo {author}
  {\bibfnamefont {P.}~\bibnamefont {Cappellaro}}, \bibinfo {author}
  {\bibfnamefont {L.}~\bibnamefont {Jiang}}, \bibinfo {author} {\bibfnamefont
  {M.~V.~G.}\ \bibnamefont {Dutt}}, \bibinfo {author} {\bibfnamefont
  {E.}~\bibnamefont {Togan}}, \bibinfo {author} {\bibfnamefont {A.~S.}\
  \bibnamefont {Zibrov}}, \bibinfo {author} {\bibfnamefont {A.}~\bibnamefont
  {Yacoby}}, \bibinfo {author} {\bibfnamefont {R.~L.}\ \bibnamefont
  {Walsworth}},\ and\ \bibinfo {author} {\bibfnamefont {M.~D.}\ \bibnamefont
  {Lukin}},\ }\bibfield  {title} {\bibinfo {title} {Nanoscale magnetic sensing
  with an individual electronic spin in diamond},\ }\href
  {https://doi.org/10.1038/nature07279} {\bibfield  {journal} {\bibinfo
  {journal} {Nature}\ }\textbf {\bibinfo {volume} {455}},\ \bibinfo {pages}
  {644} (\bibinfo {year} {2008})}\BibitemShut {NoStop}%
\bibitem [{\citenamefont {Rondin}\ \emph {et~al.}(2014)\citenamefont {Rondin},
  \citenamefont {Tetienne}, \citenamefont {Hingant}, \citenamefont {Roch},
  \citenamefont {Maletinsky},\ and\ \citenamefont
  {Jacques}}]{rondinMagnetometryNitrogenvacancyDefects2014}%
  \BibitemOpen
  \bibfield  {author} {\bibinfo {author} {\bibfnamefont {L.}~\bibnamefont
  {Rondin}}, \bibinfo {author} {\bibfnamefont {J.-P.}\ \bibnamefont
  {Tetienne}}, \bibinfo {author} {\bibfnamefont {T.}~\bibnamefont {Hingant}},
  \bibinfo {author} {\bibfnamefont {J.-F.}\ \bibnamefont {Roch}}, \bibinfo
  {author} {\bibfnamefont {P.}~\bibnamefont {Maletinsky}},\ and\ \bibinfo
  {author} {\bibfnamefont {V.}~\bibnamefont {Jacques}},\ }\bibfield  {title}
  {\bibinfo {title} {Magnetometry with nitrogen-vacancy defects in diamond},\
  }\href {https://doi.org/10.1088/0034-4885/77/5/056503} {\bibfield  {journal}
  {\bibinfo  {journal} {Reports on Progress in Physics}\ }\textbf {\bibinfo
  {volume} {77}},\ \bibinfo {pages} {056503} (\bibinfo {year}
  {2014})}\BibitemShut {NoStop}%
\bibitem [{\citenamefont {Degen}\ \emph {et~al.}(2017)\citenamefont {Degen},
  \citenamefont {Reinhard},\ and\ \citenamefont
  {Cappellaro}}]{degenQuantumSensing2017}%
  \BibitemOpen
  \bibfield  {author} {\bibinfo {author} {\bibfnamefont {C.~L.}\ \bibnamefont
  {Degen}}, \bibinfo {author} {\bibfnamefont {F.}~\bibnamefont {Reinhard}},\
  and\ \bibinfo {author} {\bibfnamefont {P.}~\bibnamefont {Cappellaro}},\
  }\bibfield  {title} {\bibinfo {title} {Quantum sensing},\ }\href
  {https://doi.org/10.1103/RevModPhys.89.035002} {\bibfield  {journal}
  {\bibinfo  {journal} {Reviews of Modern Physics}\ }\textbf {\bibinfo {volume}
  {89}},\ \bibinfo {pages} {035002} (\bibinfo {year} {2017})}\BibitemShut
  {NoStop}%
\bibitem [{\citenamefont {Barry}\ \emph {et~al.}(2020)\citenamefont {Barry},
  \citenamefont {Schloss}, \citenamefont {Bauch}, \citenamefont {Turner},
  \citenamefont {Hart}, \citenamefont {Pham},\ and\ \citenamefont
  {Walsworth}}]{barrySensitivityOptimizationNVdiamond2020}%
  \BibitemOpen
  \bibfield  {author} {\bibinfo {author} {\bibfnamefont {J.~F.}\ \bibnamefont
  {Barry}}, \bibinfo {author} {\bibfnamefont {J.~M.}\ \bibnamefont {Schloss}},
  \bibinfo {author} {\bibfnamefont {E.}~\bibnamefont {Bauch}}, \bibinfo
  {author} {\bibfnamefont {M.~J.}\ \bibnamefont {Turner}}, \bibinfo {author}
  {\bibfnamefont {C.~A.}\ \bibnamefont {Hart}}, \bibinfo {author}
  {\bibfnamefont {L.~M.}\ \bibnamefont {Pham}},\ and\ \bibinfo {author}
  {\bibfnamefont {R.~L.}\ \bibnamefont {Walsworth}},\ }\bibfield  {title}
  {\bibinfo {title} {Sensitivity optimization for {{NV-diamond}}
  magnetometry},\ }\href {https://doi.org/10.1103/RevModPhys.92.015004}
  {\bibfield  {journal} {\bibinfo  {journal} {Reviews of Modern Physics}\
  }\textbf {\bibinfo {volume} {92}},\ \bibinfo {pages} {015004} (\bibinfo
  {year} {2020})}\BibitemShut {NoStop}%
\bibitem [{\citenamefont {Pham}\ \emph {et~al.}(2012)\citenamefont {Pham},
  \citenamefont {{Bar-Gill}}, \citenamefont {Le~Sage}, \citenamefont
  {Belthangady}, \citenamefont {Stacey}, \citenamefont {Markham}, \citenamefont
  {Twitchen}, \citenamefont {Lukin},\ and\ \citenamefont
  {Walsworth}}]{phamEnhancedMetrologyUsing2012}%
  \BibitemOpen
  \bibfield  {author} {\bibinfo {author} {\bibfnamefont {L.~M.}\ \bibnamefont
  {Pham}}, \bibinfo {author} {\bibfnamefont {N.}~\bibnamefont {{Bar-Gill}}},
  \bibinfo {author} {\bibfnamefont {D.}~\bibnamefont {Le~Sage}}, \bibinfo
  {author} {\bibfnamefont {C.}~\bibnamefont {Belthangady}}, \bibinfo {author}
  {\bibfnamefont {A.}~\bibnamefont {Stacey}}, \bibinfo {author} {\bibfnamefont
  {M.}~\bibnamefont {Markham}}, \bibinfo {author} {\bibfnamefont {D.~J.}\
  \bibnamefont {Twitchen}}, \bibinfo {author} {\bibfnamefont {M.~D.}\
  \bibnamefont {Lukin}},\ and\ \bibinfo {author} {\bibfnamefont {R.~L.}\
  \bibnamefont {Walsworth}},\ }\bibfield  {title} {\bibinfo {title} {Enhanced
  metrology using preferential orientation of nitrogen-vacancy centers in
  diamond},\ }\href {https://doi.org/10.1103/PhysRevB.86.121202} {\bibfield
  {journal} {\bibinfo  {journal} {Physical Review B}\ }\textbf {\bibinfo
  {volume} {86}},\ \bibinfo {pages} {121202} (\bibinfo {year}
  {2012})}\BibitemShut {NoStop}%
\bibitem [{\citenamefont {Wrachtrup}\ and\ \citenamefont
  {Jelezko}(2006)}]{wrachtrupProcessingQuantumInformation2006}%
  \BibitemOpen
  \bibfield  {author} {\bibinfo {author} {\bibfnamefont {J.}~\bibnamefont
  {Wrachtrup}}\ and\ \bibinfo {author} {\bibfnamefont {F.}~\bibnamefont
  {Jelezko}},\ }\bibfield  {title} {\bibinfo {title} {Processing quantum
  information in diamond},\ }\href
  {https://doi.org/10.1088/0953-8984/18/21/S08} {\bibfield  {journal} {\bibinfo
   {journal} {Journal of Physics: Condensed Matter}\ }\textbf {\bibinfo
  {volume} {18}},\ \bibinfo {pages} {S807} (\bibinfo {year}
  {2006})}\BibitemShut {NoStop}%
\bibitem [{\citenamefont {Koehl}\ \emph {et~al.}(2011)\citenamefont {Koehl},
  \citenamefont {Buckley}, \citenamefont {Heremans}, \citenamefont {Calusine},\
  and\ \citenamefont {Awschalom}}]{koehlRoomTemperatureCoherent2011}%
  \BibitemOpen
  \bibfield  {author} {\bibinfo {author} {\bibfnamefont {W.~F.}\ \bibnamefont
  {Koehl}}, \bibinfo {author} {\bibfnamefont {B.~B.}\ \bibnamefont {Buckley}},
  \bibinfo {author} {\bibfnamefont {F.~J.}\ \bibnamefont {Heremans}}, \bibinfo
  {author} {\bibfnamefont {G.}~\bibnamefont {Calusine}},\ and\ \bibinfo
  {author} {\bibfnamefont {D.~D.}\ \bibnamefont {Awschalom}},\ }\bibfield
  {title} {\bibinfo {title} {Room temperature coherent control of defect spin
  qubits in silicon carbide},\ }\href {https://doi.org/10.1038/nature10562}
  {\bibfield  {journal} {\bibinfo  {journal} {Nature}\ }\textbf {\bibinfo
  {volume} {479}},\ \bibinfo {pages} {84} (\bibinfo {year} {2011})}\BibitemShut
  {NoStop}%
\bibitem [{\citenamefont {Widmann}\ \emph {et~al.}(2015)\citenamefont
  {Widmann}, \citenamefont {Lee}, \citenamefont {Rendler}, \citenamefont {Son},
  \citenamefont {Fedder}, \citenamefont {Paik}, \citenamefont {Yang},
  \citenamefont {Zhao}, \citenamefont {Yang}, \citenamefont {Booker},
  \citenamefont {Denisenko}, \citenamefont {Jamali}, \citenamefont
  {Momenzadeh}, \citenamefont {Gerhardt}, \citenamefont {Ohshima},
  \citenamefont {Gali}, \citenamefont {Janz{\'e}n},\ and\ \citenamefont
  {Wrachtrup}}]{widmannCoherentControlSingle2015}%
  \BibitemOpen
  \bibfield  {author} {\bibinfo {author} {\bibfnamefont {M.}~\bibnamefont
  {Widmann}}, \bibinfo {author} {\bibfnamefont {S.-Y.}\ \bibnamefont {Lee}},
  \bibinfo {author} {\bibfnamefont {T.}~\bibnamefont {Rendler}}, \bibinfo
  {author} {\bibfnamefont {N.~T.}\ \bibnamefont {Son}}, \bibinfo {author}
  {\bibfnamefont {H.}~\bibnamefont {Fedder}}, \bibinfo {author} {\bibfnamefont
  {S.}~\bibnamefont {Paik}}, \bibinfo {author} {\bibfnamefont {L.-P.}\
  \bibnamefont {Yang}}, \bibinfo {author} {\bibfnamefont {N.}~\bibnamefont
  {Zhao}}, \bibinfo {author} {\bibfnamefont {S.}~\bibnamefont {Yang}}, \bibinfo
  {author} {\bibfnamefont {I.}~\bibnamefont {Booker}}, \bibinfo {author}
  {\bibfnamefont {A.}~\bibnamefont {Denisenko}}, \bibinfo {author}
  {\bibfnamefont {M.}~\bibnamefont {Jamali}}, \bibinfo {author} {\bibfnamefont
  {S.~A.}\ \bibnamefont {Momenzadeh}}, \bibinfo {author} {\bibfnamefont
  {I.}~\bibnamefont {Gerhardt}}, \bibinfo {author} {\bibfnamefont
  {T.}~\bibnamefont {Ohshima}}, \bibinfo {author} {\bibfnamefont
  {A.}~\bibnamefont {Gali}}, \bibinfo {author} {\bibfnamefont {E.}~\bibnamefont
  {Janz{\'e}n}},\ and\ \bibinfo {author} {\bibfnamefont {J.}~\bibnamefont
  {Wrachtrup}},\ }\bibfield  {title} {\bibinfo {title} {Coherent control of
  single spins in silicon carbide at room temperature},\ }\href
  {https://doi.org/10.1038/nmat4145} {\bibfield  {journal} {\bibinfo  {journal}
  {Nature Materials}\ }\textbf {\bibinfo {volume} {14}},\ \bibinfo {pages}
  {164} (\bibinfo {year} {2015})}\BibitemShut {NoStop}%
\bibitem [{\citenamefont {Wang}\ \emph {et~al.}(2020)\citenamefont {Wang},
  \citenamefont {Yan}, \citenamefont {Li}, \citenamefont {Liu}, \citenamefont
  {Liu}, \citenamefont {Guo}, \citenamefont {Guo}, \citenamefont {Zhou},
  \citenamefont {Cui}, \citenamefont {Wang}, \citenamefont {Zhou},
  \citenamefont {Xu}, \citenamefont {Xu}, \citenamefont {Li},\ and\
  \citenamefont {Guo}}]{wangCoherentControlNitrogenVacancy2020}%
  \BibitemOpen
  \bibfield  {author} {\bibinfo {author} {\bibfnamefont {J.-F.}\ \bibnamefont
  {Wang}}, \bibinfo {author} {\bibfnamefont {F.-F.}\ \bibnamefont {Yan}},
  \bibinfo {author} {\bibfnamefont {Q.}~\bibnamefont {Li}}, \bibinfo {author}
  {\bibfnamefont {Z.-H.}\ \bibnamefont {Liu}}, \bibinfo {author} {\bibfnamefont
  {H.}~\bibnamefont {Liu}}, \bibinfo {author} {\bibfnamefont {G.-P.}\
  \bibnamefont {Guo}}, \bibinfo {author} {\bibfnamefont {L.-P.}\ \bibnamefont
  {Guo}}, \bibinfo {author} {\bibfnamefont {X.}~\bibnamefont {Zhou}}, \bibinfo
  {author} {\bibfnamefont {J.-M.}\ \bibnamefont {Cui}}, \bibinfo {author}
  {\bibfnamefont {J.}~\bibnamefont {Wang}}, \bibinfo {author} {\bibfnamefont
  {Z.-Q.}\ \bibnamefont {Zhou}}, \bibinfo {author} {\bibfnamefont {X.-Y.}\
  \bibnamefont {Xu}}, \bibinfo {author} {\bibfnamefont {J.-S.}\ \bibnamefont
  {Xu}}, \bibinfo {author} {\bibfnamefont {C.-F.}\ \bibnamefont {Li}},\ and\
  \bibinfo {author} {\bibfnamefont {G.-C.}\ \bibnamefont {Guo}},\ }\bibfield
  {title} {\bibinfo {title} {Coherent {{Control}} of {{Nitrogen-Vacancy Center
  Spins}} in {{Silicon Carbide}} at {{Room Temperature}}},\ }\href
  {https://doi.org/10.1103/PhysRevLett.124.223601} {\bibfield  {journal}
  {\bibinfo  {journal} {Physical Review Letters}\ }\textbf {\bibinfo {volume}
  {124}},\ \bibinfo {pages} {223601} (\bibinfo {year} {2020})}\BibitemShut
  {NoStop}%
\bibitem [{\citenamefont {Son}\ \emph {et~al.}(2020)\citenamefont {Son},
  \citenamefont {Anderson}, \citenamefont {Bourassa}, \citenamefont {Miao},
  \citenamefont {Babin}, \citenamefont {Widmann}, \citenamefont {Niethammer},
  \citenamefont {Ul~Hassan}, \citenamefont {Morioka}, \citenamefont {Ivanov},
  \citenamefont {Kaiser}, \citenamefont {Wrachtrup},\ and\ \citenamefont
  {Awschalom}}]{sonDevelopingSiliconCarbide2020}%
  \BibitemOpen
  \bibfield  {author} {\bibinfo {author} {\bibfnamefont {N.~T.}\ \bibnamefont
  {Son}}, \bibinfo {author} {\bibfnamefont {C.~P.}\ \bibnamefont {Anderson}},
  \bibinfo {author} {\bibfnamefont {A.}~\bibnamefont {Bourassa}}, \bibinfo
  {author} {\bibfnamefont {K.~C.}\ \bibnamefont {Miao}}, \bibinfo {author}
  {\bibfnamefont {C.}~\bibnamefont {Babin}}, \bibinfo {author} {\bibfnamefont
  {M.}~\bibnamefont {Widmann}}, \bibinfo {author} {\bibfnamefont
  {M.}~\bibnamefont {Niethammer}}, \bibinfo {author} {\bibfnamefont
  {J.}~\bibnamefont {Ul~Hassan}}, \bibinfo {author} {\bibfnamefont
  {N.}~\bibnamefont {Morioka}}, \bibinfo {author} {\bibfnamefont {I.~G.}\
  \bibnamefont {Ivanov}}, \bibinfo {author} {\bibfnamefont {F.}~\bibnamefont
  {Kaiser}}, \bibinfo {author} {\bibfnamefont {J.}~\bibnamefont {Wrachtrup}},\
  and\ \bibinfo {author} {\bibfnamefont {D.~D.}\ \bibnamefont {Awschalom}},\
  }\bibfield  {title} {\bibinfo {title} {Developing silicon carbide for quantum
  spintronics},\ }\href {https://doi.org/10.1063/5.0004454} {\bibfield
  {journal} {\bibinfo  {journal} {Applied Physics Letters}\ }\textbf {\bibinfo
  {volume} {116}},\ \bibinfo {pages} {190501} (\bibinfo {year}
  {2020})}\BibitemShut {NoStop}%
\bibitem [{\citenamefont {Castelletto}\ \emph {et~al.}(2023)\citenamefont
  {Castelletto}, \citenamefont {Lew}, \citenamefont {Lin},\ and\ \citenamefont
  {Xu}}]{castellettoQuantumSystemsSilicon2023}%
  \BibitemOpen
  \bibfield  {author} {\bibinfo {author} {\bibfnamefont {S.}~\bibnamefont
  {Castelletto}}, \bibinfo {author} {\bibfnamefont {C.~T.-K.}\ \bibnamefont
  {Lew}}, \bibinfo {author} {\bibfnamefont {W.-X.}\ \bibnamefont {Lin}},\ and\
  \bibinfo {author} {\bibfnamefont {J.-S.}\ \bibnamefont {Xu}},\ }\bibfield
  {title} {\bibinfo {title} {Quantum systems in silicon carbide for sensing
  applications},\ }\href {https://doi.org/10.1088/1361-6633/ad10b3} {\bibfield
  {journal} {\bibinfo  {journal} {Reports on Progress in Physics}\ }\textbf
  {\bibinfo {volume} {87}},\ \bibinfo {pages} {014501} (\bibinfo {year}
  {2023})}\BibitemShut {NoStop}%
\bibitem [{\citenamefont {Carter}\ \emph {et~al.}(2015)\citenamefont {Carter},
  \citenamefont {Soykal}, \citenamefont {Dev}, \citenamefont {Economou},\ and\
  \citenamefont {Glaser}}]{carterSpinCoherenceEcho2015}%
  \BibitemOpen
  \bibfield  {author} {\bibinfo {author} {\bibfnamefont {S.~G.}\ \bibnamefont
  {Carter}}, \bibinfo {author} {\bibfnamefont {{\"O}.~O.}\ \bibnamefont
  {Soykal}}, \bibinfo {author} {\bibfnamefont {P.}~\bibnamefont {Dev}},
  \bibinfo {author} {\bibfnamefont {S.~E.}\ \bibnamefont {Economou}},\ and\
  \bibinfo {author} {\bibfnamefont {E.~R.}\ \bibnamefont {Glaser}},\ }\bibfield
   {title} {\bibinfo {title} {Spin coherence and echo modulation of the silicon
  vacancy in {{4H-SiC}} at room temperature},\ }\href
  {https://doi.org/10.1103/PhysRevB.92.161202} {\bibfield  {journal} {\bibinfo
  {journal} {Physical Review B}\ }\textbf {\bibinfo {volume} {92}},\ \bibinfo
  {pages} {161202} (\bibinfo {year} {2015})}\BibitemShut {NoStop}%
\bibitem [{\citenamefont {Tarasenko}\ \emph {et~al.}(2018)\citenamefont
  {Tarasenko}, \citenamefont {Poshakinskiy}, \citenamefont {Simin},
  \citenamefont {Soltamov}, \citenamefont {Mokhov}, \citenamefont {Baranov},
  \citenamefont {Dyakonov},\ and\ \citenamefont
  {Astakhov}}]{tarasenkoSpinOpticalProperties2018}%
  \BibitemOpen
  \bibfield  {author} {\bibinfo {author} {\bibfnamefont {S.~A.}\ \bibnamefont
  {Tarasenko}}, \bibinfo {author} {\bibfnamefont {A.~V.}\ \bibnamefont
  {Poshakinskiy}}, \bibinfo {author} {\bibfnamefont {D.}~\bibnamefont {Simin}},
  \bibinfo {author} {\bibfnamefont {V.~A.}\ \bibnamefont {Soltamov}}, \bibinfo
  {author} {\bibfnamefont {E.~N.}\ \bibnamefont {Mokhov}}, \bibinfo {author}
  {\bibfnamefont {P.~G.}\ \bibnamefont {Baranov}}, \bibinfo {author}
  {\bibfnamefont {V.}~\bibnamefont {Dyakonov}},\ and\ \bibinfo {author}
  {\bibfnamefont {G.~V.}\ \bibnamefont {Astakhov}},\ }\bibfield  {title}
  {\bibinfo {title} {Spin and {{Optical Properties}} of {{Silicon Vacancies}}
  in {{Silicon Carbide}} - {{A Review}}},\ }\href
  {https://doi.org/10.1002/pssb.201700258} {\bibfield  {journal} {\bibinfo
  {journal} {physica status solidi (b)}\ }\textbf {\bibinfo {volume} {255}},\
  \bibinfo {pages} {1700258} (\bibinfo {year} {2018})}\BibitemShut {NoStop}%
\bibitem [{\citenamefont {Iv{\'a}dy}\ \emph {et~al.}(2017)\citenamefont
  {Iv{\'a}dy}, \citenamefont {Davidsson}, \citenamefont {Son}, \citenamefont
  {Ohshima}, \citenamefont {Abrikosov},\ and\ \citenamefont
  {Gali}}]{ivadyIdentificationSivacancyRelated2017}%
  \BibitemOpen
  \bibfield  {author} {\bibinfo {author} {\bibfnamefont {V.}~\bibnamefont
  {Iv{\'a}dy}}, \bibinfo {author} {\bibfnamefont {J.}~\bibnamefont
  {Davidsson}}, \bibinfo {author} {\bibfnamefont {N.~T.}\ \bibnamefont {Son}},
  \bibinfo {author} {\bibfnamefont {T.}~\bibnamefont {Ohshima}}, \bibinfo
  {author} {\bibfnamefont {I.~A.}\ \bibnamefont {Abrikosov}},\ and\ \bibinfo
  {author} {\bibfnamefont {A.}~\bibnamefont {Gali}},\ }\bibfield  {title}
  {\bibinfo {title} {Identification of {{Si-vacancy}} related room-temperature
  qubits in {{4H}} silicon carbide},\ }\href
  {https://doi.org/10.1103/PhysRevB.96.161114} {\bibfield  {journal} {\bibinfo
  {journal} {Physical Review B}\ }\textbf {\bibinfo {volume} {96}},\ \bibinfo
  {pages} {161114} (\bibinfo {year} {2017})}\BibitemShut {NoStop}%
\bibitem [{\citenamefont {Liu}\ \emph {et~al.}(2024)\citenamefont {Liu},
  \citenamefont {Kaiser}, \citenamefont {Bushmakin}, \citenamefont
  {Hesselmeier}, \citenamefont {Steidl}, \citenamefont {Ohshima}, \citenamefont
  {Son}, \citenamefont {{Ul-Hassan}}, \citenamefont {Soykal},\ and\
  \citenamefont {Wrachtrup}}]{liuSiliconVacancycenters2024}%
  \BibitemOpen
  \bibfield  {author} {\bibinfo {author} {\bibfnamefont {D.}~\bibnamefont
  {Liu}}, \bibinfo {author} {\bibfnamefont {F.}~\bibnamefont {Kaiser}},
  \bibinfo {author} {\bibfnamefont {V.}~\bibnamefont {Bushmakin}}, \bibinfo
  {author} {\bibfnamefont {E.}~\bibnamefont {Hesselmeier}}, \bibinfo {author}
  {\bibfnamefont {T.}~\bibnamefont {Steidl}}, \bibinfo {author} {\bibfnamefont
  {T.}~\bibnamefont {Ohshima}}, \bibinfo {author} {\bibfnamefont {N.~T.}\
  \bibnamefont {Son}}, \bibinfo {author} {\bibfnamefont {J.}~\bibnamefont
  {{Ul-Hassan}}}, \bibinfo {author} {\bibfnamefont {{\"O}.~O.}\ \bibnamefont
  {Soykal}},\ and\ \bibinfo {author} {\bibfnamefont {J.}~\bibnamefont
  {Wrachtrup}},\ }\bibfield  {title} {\bibinfo {title} {The silicon vacancy
  centers in {{SiC}}: Determination of intrinsic spin dynamics for integrated
  quantum photonics},\ }\href {https://doi.org/10.1038/s41534-024-00861-6}
  {\bibfield  {journal} {\bibinfo  {journal} {npj Quantum Information}\
  }\textbf {\bibinfo {volume} {10}},\ \bibinfo {pages} {72} (\bibinfo {year}
  {2024})}\BibitemShut {NoStop}%
\bibitem [{\citenamefont {Kraus}\ \emph {et~al.}(2017)\citenamefont {Kraus},
  \citenamefont {Simin}, \citenamefont {Kasper}, \citenamefont {Suda},
  \citenamefont {Kawabata}, \citenamefont {Kada}, \citenamefont {Honda},
  \citenamefont {Hijikata}, \citenamefont {Ohshima}, \citenamefont {Dyakonov},\
  and\ \citenamefont {Astakhov}}]{krausThreeDimensionalProtonBeam2017}%
  \BibitemOpen
  \bibfield  {author} {\bibinfo {author} {\bibfnamefont {H.}~\bibnamefont
  {Kraus}}, \bibinfo {author} {\bibfnamefont {D.}~\bibnamefont {Simin}},
  \bibinfo {author} {\bibfnamefont {C.}~\bibnamefont {Kasper}}, \bibinfo
  {author} {\bibfnamefont {Y.}~\bibnamefont {Suda}}, \bibinfo {author}
  {\bibfnamefont {S.}~\bibnamefont {Kawabata}}, \bibinfo {author}
  {\bibfnamefont {W.}~\bibnamefont {Kada}}, \bibinfo {author} {\bibfnamefont
  {T.}~\bibnamefont {Honda}}, \bibinfo {author} {\bibfnamefont
  {Y.}~\bibnamefont {Hijikata}}, \bibinfo {author} {\bibfnamefont
  {T.}~\bibnamefont {Ohshima}}, \bibinfo {author} {\bibfnamefont
  {V.}~\bibnamefont {Dyakonov}},\ and\ \bibinfo {author} {\bibfnamefont
  {G.~V.}\ \bibnamefont {Astakhov}},\ }\bibfield  {title} {\bibinfo {title}
  {Three-{{Dimensional Proton Beam Writing}} of {{Optically Active Coherent
  Vacancy Spins}} in {{Silicon Carbide}}},\ }\href
  {https://doi.org/10.1021/acs.nanolett.6b05395} {\bibfield  {journal}
  {\bibinfo  {journal} {Nano Letters}\ }\textbf {\bibinfo {volume} {17}},\
  \bibinfo {pages} {2865} (\bibinfo {year} {2017})}\BibitemShut {NoStop}%
\bibitem [{\citenamefont {Ohshima}\ \emph {et~al.}(2018)\citenamefont
  {Ohshima}, \citenamefont {Satoh}, \citenamefont {Kraus}, \citenamefont
  {Astakhov}, \citenamefont {Dyakonov},\ and\ \citenamefont
  {Baranov}}]{ohshimaCreationSiliconVacancy2018}%
  \BibitemOpen
  \bibfield  {author} {\bibinfo {author} {\bibfnamefont {T.}~\bibnamefont
  {Ohshima}}, \bibinfo {author} {\bibfnamefont {T.}~\bibnamefont {Satoh}},
  \bibinfo {author} {\bibfnamefont {H.}~\bibnamefont {Kraus}}, \bibinfo
  {author} {\bibfnamefont {G.~V.}\ \bibnamefont {Astakhov}}, \bibinfo {author}
  {\bibfnamefont {V.}~\bibnamefont {Dyakonov}},\ and\ \bibinfo {author}
  {\bibfnamefont {P.~G.}\ \bibnamefont {Baranov}},\ }\bibfield  {title}
  {\bibinfo {title} {Creation of silicon vacancy in silicon carbide by proton
  beam writing toward quantum sensing applications},\ }\href
  {https://doi.org/10.1088/1361-6463/aad0ec} {\bibfield  {journal} {\bibinfo
  {journal} {Journal of Physics D: Applied Physics}\ }\textbf {\bibinfo
  {volume} {51}},\ \bibinfo {pages} {333002} (\bibinfo {year}
  {2018})}\BibitemShut {NoStop}%
\bibitem [{\citenamefont {Kasper}\ \emph {et~al.}(2020)\citenamefont {Kasper},
  \citenamefont {Klenkert}, \citenamefont {Shang}, \citenamefont {Simin},
  \citenamefont {Gottscholl}, \citenamefont {Sperlich}, \citenamefont {Kraus},
  \citenamefont {Schneider}, \citenamefont {Zhou}, \citenamefont {Trupke},
  \citenamefont {Kada}, \citenamefont {Ohshima}, \citenamefont {Dyakonov},\
  and\ \citenamefont {Astakhov}}]{kasperInfluenceIrradiationDefect2020}%
  \BibitemOpen
  \bibfield  {author} {\bibinfo {author} {\bibfnamefont {C.}~\bibnamefont
  {Kasper}}, \bibinfo {author} {\bibfnamefont {D.}~\bibnamefont {Klenkert}},
  \bibinfo {author} {\bibfnamefont {Z.}~\bibnamefont {Shang}}, \bibinfo
  {author} {\bibfnamefont {D.}~\bibnamefont {Simin}}, \bibinfo {author}
  {\bibfnamefont {A.}~\bibnamefont {Gottscholl}}, \bibinfo {author}
  {\bibfnamefont {A.}~\bibnamefont {Sperlich}}, \bibinfo {author}
  {\bibfnamefont {H.}~\bibnamefont {Kraus}}, \bibinfo {author} {\bibfnamefont
  {C.}~\bibnamefont {Schneider}}, \bibinfo {author} {\bibfnamefont
  {S.}~\bibnamefont {Zhou}}, \bibinfo {author} {\bibfnamefont {M.}~\bibnamefont
  {Trupke}}, \bibinfo {author} {\bibfnamefont {W.}~\bibnamefont {Kada}},
  \bibinfo {author} {\bibfnamefont {T.}~\bibnamefont {Ohshima}}, \bibinfo
  {author} {\bibfnamefont {V.}~\bibnamefont {Dyakonov}},\ and\ \bibinfo
  {author} {\bibfnamefont {G.~V.}\ \bibnamefont {Astakhov}},\ }\bibfield
  {title} {\bibinfo {title} {Influence of {{Irradiation}} on {{Defect Spin
  Coherence}} in {{Silicon Carbide}}},\ }\href
  {https://doi.org/10.1103/PhysRevApplied.13.044054} {\bibfield  {journal}
  {\bibinfo  {journal} {Physical Review Applied}\ }\textbf {\bibinfo {volume}
  {13}},\ \bibinfo {pages} {044054} (\bibinfo {year} {2020})}\BibitemShut
  {NoStop}%
\bibitem [{\citenamefont {Niethammer}\ \emph {et~al.}(2019)\citenamefont
  {Niethammer}, \citenamefont {Widmann}, \citenamefont {Rendler}, \citenamefont
  {Morioka}, \citenamefont {Chen}, \citenamefont {St{\"o}hr}, \citenamefont
  {Hassan}, \citenamefont {Onoda}, \citenamefont {Ohshima}, \citenamefont
  {Lee}, \citenamefont {Mukherjee}, \citenamefont {Isoya}, \citenamefont
  {Son},\ and\ \citenamefont
  {Wrachtrup}}]{niethammerCoherentElectricalReadout2019}%
  \BibitemOpen
  \bibfield  {author} {\bibinfo {author} {\bibfnamefont {M.}~\bibnamefont
  {Niethammer}}, \bibinfo {author} {\bibfnamefont {M.}~\bibnamefont {Widmann}},
  \bibinfo {author} {\bibfnamefont {T.}~\bibnamefont {Rendler}}, \bibinfo
  {author} {\bibfnamefont {N.}~\bibnamefont {Morioka}}, \bibinfo {author}
  {\bibfnamefont {Y.-C.}\ \bibnamefont {Chen}}, \bibinfo {author}
  {\bibfnamefont {R.}~\bibnamefont {St{\"o}hr}}, \bibinfo {author}
  {\bibfnamefont {J.~U.}\ \bibnamefont {Hassan}}, \bibinfo {author}
  {\bibfnamefont {S.}~\bibnamefont {Onoda}}, \bibinfo {author} {\bibfnamefont
  {T.}~\bibnamefont {Ohshima}}, \bibinfo {author} {\bibfnamefont {S.-Y.}\
  \bibnamefont {Lee}}, \bibinfo {author} {\bibfnamefont {A.}~\bibnamefont
  {Mukherjee}}, \bibinfo {author} {\bibfnamefont {J.}~\bibnamefont {Isoya}},
  \bibinfo {author} {\bibfnamefont {N.~T.}\ \bibnamefont {Son}},\ and\ \bibinfo
  {author} {\bibfnamefont {J.}~\bibnamefont {Wrachtrup}},\ }\bibfield  {title}
  {\bibinfo {title} {Coherent electrical readout of defect spins in silicon
  carbide by photo-ionization at ambient conditions},\ }\href
  {https://doi.org/10.1038/s41467-019-13545-z} {\bibfield  {journal} {\bibinfo
  {journal} {Nature Communications}\ }\textbf {\bibinfo {volume} {10}},\
  \bibinfo {pages} {5569} (\bibinfo {year} {2019})}\BibitemShut {NoStop}%
\bibitem [{\citenamefont {Hoang}\ \emph {et~al.}(2021)\citenamefont {Hoang},
  \citenamefont {Ishiwata}, \citenamefont {Masuyama}, \citenamefont {Yamazaki},
  \citenamefont {Kojima}, \citenamefont {Lee}, \citenamefont {Ohshima},
  \citenamefont {Iwasaki}, \citenamefont {Hisamoto},\ and\ \citenamefont
  {Hatano}}]{hoangThermometricQuantumSensor2021}%
  \BibitemOpen
  \bibfield  {author} {\bibinfo {author} {\bibfnamefont {T.~M.}\ \bibnamefont
  {Hoang}}, \bibinfo {author} {\bibfnamefont {H.}~\bibnamefont {Ishiwata}},
  \bibinfo {author} {\bibfnamefont {Y.}~\bibnamefont {Masuyama}}, \bibinfo
  {author} {\bibfnamefont {Y.}~\bibnamefont {Yamazaki}}, \bibinfo {author}
  {\bibfnamefont {K.}~\bibnamefont {Kojima}}, \bibinfo {author} {\bibfnamefont
  {S.-Y.}\ \bibnamefont {Lee}}, \bibinfo {author} {\bibfnamefont
  {T.}~\bibnamefont {Ohshima}}, \bibinfo {author} {\bibfnamefont
  {T.}~\bibnamefont {Iwasaki}}, \bibinfo {author} {\bibfnamefont
  {D.}~\bibnamefont {Hisamoto}},\ and\ \bibinfo {author} {\bibfnamefont
  {M.}~\bibnamefont {Hatano}},\ }\bibfield  {title} {\bibinfo {title}
  {Thermometric quantum sensor using excited state of silicon vacancy centers
  in {{4H-SiC}} devices},\ }\href {https://doi.org/10.1063/5.0027603}
  {\bibfield  {journal} {\bibinfo  {journal} {Applied Physics Letters}\
  }\textbf {\bibinfo {volume} {118}},\ \bibinfo {pages} {044001} (\bibinfo
  {year} {2021})}\BibitemShut {NoStop}%
\bibitem [{\citenamefont {Li}\ \emph {et~al.}(2022)\citenamefont {Li},
  \citenamefont {Wang}, \citenamefont {Yan}, \citenamefont {Zhou},
  \citenamefont {Wang}, \citenamefont {Liu}, \citenamefont {Guo}, \citenamefont
  {Zhou}, \citenamefont {Gali}, \citenamefont {Liu}, \citenamefont {Wang},
  \citenamefont {Sun}, \citenamefont {Guo}, \citenamefont {Tang}, \citenamefont
  {Li}, \citenamefont {You}, \citenamefont {Xu}, \citenamefont {Li},\ and\
  \citenamefont {Guo}}]{liRoomtemperatureCoherentManipulation2022}%
  \BibitemOpen
  \bibfield  {author} {\bibinfo {author} {\bibfnamefont {Q.}~\bibnamefont
  {Li}}, \bibinfo {author} {\bibfnamefont {J.-F.}\ \bibnamefont {Wang}},
  \bibinfo {author} {\bibfnamefont {F.-F.}\ \bibnamefont {Yan}}, \bibinfo
  {author} {\bibfnamefont {J.-Y.}\ \bibnamefont {Zhou}}, \bibinfo {author}
  {\bibfnamefont {H.-F.}\ \bibnamefont {Wang}}, \bibinfo {author}
  {\bibfnamefont {H.}~\bibnamefont {Liu}}, \bibinfo {author} {\bibfnamefont
  {L.-P.}\ \bibnamefont {Guo}}, \bibinfo {author} {\bibfnamefont
  {X.}~\bibnamefont {Zhou}}, \bibinfo {author} {\bibfnamefont {A.}~\bibnamefont
  {Gali}}, \bibinfo {author} {\bibfnamefont {Z.-H.}\ \bibnamefont {Liu}},
  \bibinfo {author} {\bibfnamefont {Z.-Q.}\ \bibnamefont {Wang}}, \bibinfo
  {author} {\bibfnamefont {K.}~\bibnamefont {Sun}}, \bibinfo {author}
  {\bibfnamefont {G.-P.}\ \bibnamefont {Guo}}, \bibinfo {author} {\bibfnamefont
  {J.-S.}\ \bibnamefont {Tang}}, \bibinfo {author} {\bibfnamefont
  {H.}~\bibnamefont {Li}}, \bibinfo {author} {\bibfnamefont {L.-X.}\
  \bibnamefont {You}}, \bibinfo {author} {\bibfnamefont {J.-S.}\ \bibnamefont
  {Xu}}, \bibinfo {author} {\bibfnamefont {C.-F.}\ \bibnamefont {Li}},\ and\
  \bibinfo {author} {\bibfnamefont {G.-C.}\ \bibnamefont {Guo}},\ }\bibfield
  {title} {\bibinfo {title} {Room-temperature coherent manipulation of
  single-spin qubits in silicon carbide with a high readout contrast},\ }\href
  {https://doi.org/10.1093/nsr/nwab122} {\bibfield  {journal} {\bibinfo
  {journal} {National Science Review}\ }\textbf {\bibinfo {volume} {9}},\
  \bibinfo {pages} {nwab122} (\bibinfo {year} {2022})}\BibitemShut {NoStop}%
\bibitem [{\citenamefont {Lee}\ \emph {et~al.}(2015)\citenamefont {Lee},
  \citenamefont {Niethammer},\ and\ \citenamefont
  {Wrachtrup}}]{leeVectorMagnetometryBased2015}%
  \BibitemOpen
  \bibfield  {author} {\bibinfo {author} {\bibfnamefont {S.-Y.}\ \bibnamefont
  {Lee}}, \bibinfo {author} {\bibfnamefont {M.}~\bibnamefont {Niethammer}},\
  and\ \bibinfo {author} {\bibfnamefont {J.}~\bibnamefont {Wrachtrup}},\
  }\bibfield  {title} {\bibinfo {title} {Vector magnetometry based on
  {$S=\frac{3}{2}$} electronic spins},\ }\href
  {https://doi.org/10.1103/PhysRevB.92.115201} {\bibfield  {journal} {\bibinfo
  {journal} {Physical Review B}\ }\textbf {\bibinfo {volume} {92}},\ \bibinfo
  {pages} {115201} (\bibinfo {year} {2015})}\BibitemShut {NoStop}%
\bibitem [{\citenamefont {Niethammer}\ \emph {et~al.}(2016)\citenamefont
  {Niethammer}, \citenamefont {Widmann}, \citenamefont {Lee}, \citenamefont
  {Stenberg}, \citenamefont {Kordina}, \citenamefont {Ohshima}, \citenamefont
  {Son}, \citenamefont {Janz{\'e}n},\ and\ \citenamefont
  {Wrachtrup}}]{niethammerVectorMagnetometryUsing2016}%
  \BibitemOpen
  \bibfield  {author} {\bibinfo {author} {\bibfnamefont {M.}~\bibnamefont
  {Niethammer}}, \bibinfo {author} {\bibfnamefont {M.}~\bibnamefont {Widmann}},
  \bibinfo {author} {\bibfnamefont {S.-Y.}\ \bibnamefont {Lee}}, \bibinfo
  {author} {\bibfnamefont {P.}~\bibnamefont {Stenberg}}, \bibinfo {author}
  {\bibfnamefont {O.}~\bibnamefont {Kordina}}, \bibinfo {author} {\bibfnamefont
  {T.}~\bibnamefont {Ohshima}}, \bibinfo {author} {\bibfnamefont {N.~T.}\
  \bibnamefont {Son}}, \bibinfo {author} {\bibfnamefont {E.}~\bibnamefont
  {Janz{\'e}n}},\ and\ \bibinfo {author} {\bibfnamefont {J.}~\bibnamefont
  {Wrachtrup}},\ }\bibfield  {title} {\bibinfo {title} {Vector {{Magnetometry
  Using Silicon Vacancies}} in {{4H-SiC Under Ambient Conditions}}},\ }\href
  {https://doi.org/10.1103/PhysRevApplied.6.034001} {\bibfield  {journal}
  {\bibinfo  {journal} {Physical Review Applied}\ }\textbf {\bibinfo {volume}
  {6}},\ \bibinfo {pages} {034001} (\bibinfo {year} {2016})}\BibitemShut
  {NoStop}%
\bibitem [{\citenamefont {Simin}\ \emph {et~al.}(2016)\citenamefont {Simin},
  \citenamefont {Soltamov}, \citenamefont {Poshakinskiy}, \citenamefont
  {Anisimov}, \citenamefont {Babunts}, \citenamefont {Tolmachev}, \citenamefont
  {Mokhov}, \citenamefont {Trupke}, \citenamefont {Tarasenko}, \citenamefont
  {Sperlich}, \citenamefont {Baranov}, \citenamefont {Dyakonov},\ and\
  \citenamefont {Astakhov}}]{siminAllOpticalDcNanotesla2016}%
  \BibitemOpen
  \bibfield  {author} {\bibinfo {author} {\bibfnamefont {D.}~\bibnamefont
  {Simin}}, \bibinfo {author} {\bibfnamefont {V.~A.}\ \bibnamefont {Soltamov}},
  \bibinfo {author} {\bibfnamefont {A.~V.}\ \bibnamefont {Poshakinskiy}},
  \bibinfo {author} {\bibfnamefont {A.~N.}\ \bibnamefont {Anisimov}}, \bibinfo
  {author} {\bibfnamefont {R.~A.}\ \bibnamefont {Babunts}}, \bibinfo {author}
  {\bibfnamefont {D.~O.}\ \bibnamefont {Tolmachev}}, \bibinfo {author}
  {\bibfnamefont {E.~N.}\ \bibnamefont {Mokhov}}, \bibinfo {author}
  {\bibfnamefont {M.}~\bibnamefont {Trupke}}, \bibinfo {author} {\bibfnamefont
  {S.~A.}\ \bibnamefont {Tarasenko}}, \bibinfo {author} {\bibfnamefont
  {A.}~\bibnamefont {Sperlich}}, \bibinfo {author} {\bibfnamefont {P.~G.}\
  \bibnamefont {Baranov}}, \bibinfo {author} {\bibfnamefont {V.}~\bibnamefont
  {Dyakonov}},\ and\ \bibinfo {author} {\bibfnamefont {G.~V.}\ \bibnamefont
  {Astakhov}},\ }\bibfield  {title} {\bibinfo {title} {All-{{Optical}} dc
  {{Nanotesla Magnetometry Using Silicon Vacancy Fine Structure}} in
  {{Isotopically Purified Silicon Carbide}}},\ }\href
  {https://doi.org/10.1103/PhysRevX.6.031014} {\bibfield  {journal} {\bibinfo
  {journal} {Physical Review X}\ }\textbf {\bibinfo {volume} {6}},\ \bibinfo
  {pages} {031014} (\bibinfo {year} {2016})}\BibitemShut {NoStop}%
\bibitem [{\citenamefont {Anisimov}\ \emph {et~al.}(2016)\citenamefont
  {Anisimov}, \citenamefont {Simin}, \citenamefont {Soltamov}, \citenamefont
  {Lebedev}, \citenamefont {Baranov}, \citenamefont {Astakhov},\ and\
  \citenamefont {Dyakonov}}]{anisimovOpticalThermometryBased2016}%
  \BibitemOpen
  \bibfield  {author} {\bibinfo {author} {\bibfnamefont {A.~N.}\ \bibnamefont
  {Anisimov}}, \bibinfo {author} {\bibfnamefont {D.}~\bibnamefont {Simin}},
  \bibinfo {author} {\bibfnamefont {V.~A.}\ \bibnamefont {Soltamov}}, \bibinfo
  {author} {\bibfnamefont {S.~P.}\ \bibnamefont {Lebedev}}, \bibinfo {author}
  {\bibfnamefont {P.~G.}\ \bibnamefont {Baranov}}, \bibinfo {author}
  {\bibfnamefont {G.~V.}\ \bibnamefont {Astakhov}},\ and\ \bibinfo {author}
  {\bibfnamefont {V.}~\bibnamefont {Dyakonov}},\ }\bibfield  {title} {\bibinfo
  {title} {Optical thermometry based on level anticrossing in silicon
  carbide},\ }\href {https://doi.org/10.1038/srep33301} {\bibfield  {journal}
  {\bibinfo  {journal} {Scientific Reports}\ }\textbf {\bibinfo {volume} {6}},\
  \bibinfo {pages} {33301} (\bibinfo {year} {2016})}\BibitemShut {NoStop}%
\bibitem [{\citenamefont {Soykal}\ and\ \citenamefont
  {Reinecke}(2017)}]{soykalQuantumMetrologySingle2017}%
  \BibitemOpen
  \bibfield  {author} {\bibinfo {author} {\bibfnamefont {{\"O}.~O.}\
  \bibnamefont {Soykal}}\ and\ \bibinfo {author} {\bibfnamefont {T.~L.}\
  \bibnamefont {Reinecke}},\ }\bibfield  {title} {\bibinfo {title} {Quantum
  metrology with a single spin-$\frac{3}{2}$ defect in silicon carbide},\
  }\href {https://doi.org/10.1103/PhysRevB.95.081405} {\bibfield  {journal}
  {\bibinfo  {journal} {Physical Review B}\ }\textbf {\bibinfo {volume} {95}},\
  \bibinfo {pages} {081405} (\bibinfo {year} {2017})}\BibitemShut {NoStop}%
\bibitem [{\citenamefont {Dong}\ \emph {et~al.}(2019)\citenamefont {Dong},
  \citenamefont {Doherty},\ and\ \citenamefont
  {Economou}}]{dongSpinPolarizationIntersystem2019}%
  \BibitemOpen
  \bibfield  {author} {\bibinfo {author} {\bibfnamefont {W.}~\bibnamefont
  {Dong}}, \bibinfo {author} {\bibfnamefont {M.~W.}\ \bibnamefont {Doherty}},\
  and\ \bibinfo {author} {\bibfnamefont {S.~E.}\ \bibnamefont {Economou}},\
  }\bibfield  {title} {\bibinfo {title} {Spin polarization through intersystem
  crossing in the silicon vacancy of silicon carbide},\ }\href
  {https://doi.org/10.1103/PhysRevB.99.184102} {\bibfield  {journal} {\bibinfo
  {journal} {Physical Review B}\ }\textbf {\bibinfo {volume} {99}},\ \bibinfo
  {pages} {184102} (\bibinfo {year} {2019})}\BibitemShut {NoStop}%
\bibitem [{\citenamefont {Zhang}\ \emph {et~al.}(2021)\citenamefont {Zhang},
  \citenamefont {Shagieva}, \citenamefont {Widmann}, \citenamefont
  {K{\"u}bler}, \citenamefont {Vorobyov}, \citenamefont {Kapitanova},
  \citenamefont {Nenasheva}, \citenamefont {Corkill}, \citenamefont {Rhrle},
  \citenamefont {Nakamura}, \citenamefont {Sumiya}, \citenamefont {Onoda},
  \citenamefont {Isoya},\ and\ \citenamefont
  {Wrachtrup}}]{zhangDiamondMagnetometryGradiometry2021}%
  \BibitemOpen
  \bibfield  {author} {\bibinfo {author} {\bibfnamefont {C.}~\bibnamefont
  {Zhang}}, \bibinfo {author} {\bibfnamefont {F.}~\bibnamefont {Shagieva}},
  \bibinfo {author} {\bibfnamefont {M.}~\bibnamefont {Widmann}}, \bibinfo
  {author} {\bibfnamefont {M.}~\bibnamefont {K{\"u}bler}}, \bibinfo {author}
  {\bibfnamefont {V.}~\bibnamefont {Vorobyov}}, \bibinfo {author}
  {\bibfnamefont {P.}~\bibnamefont {Kapitanova}}, \bibinfo {author}
  {\bibfnamefont {E.}~\bibnamefont {Nenasheva}}, \bibinfo {author}
  {\bibfnamefont {R.}~\bibnamefont {Corkill}}, \bibinfo {author} {\bibfnamefont
  {O.}~\bibnamefont {Rhrle}}, \bibinfo {author} {\bibfnamefont
  {K.}~\bibnamefont {Nakamura}}, \bibinfo {author} {\bibfnamefont
  {H.}~\bibnamefont {Sumiya}}, \bibinfo {author} {\bibfnamefont
  {S.}~\bibnamefont {Onoda}}, \bibinfo {author} {\bibfnamefont
  {J.}~\bibnamefont {Isoya}},\ and\ \bibinfo {author} {\bibfnamefont
  {J.}~\bibnamefont {Wrachtrup}},\ }\bibfield  {title} {\bibinfo {title}
  {Diamond {{Magnetometry}} and {{Gradiometry Towards Subpicotesla}} dc {{Field
  Measurement}}},\ }\href {https://doi.org/10.1103/PhysRevApplied.15.064075}
  {\bibfield  {journal} {\bibinfo  {journal} {Physical Review Applied}\
  }\textbf {\bibinfo {volume} {15}},\ \bibinfo {pages} {064075} (\bibinfo
  {year} {2021})}\BibitemShut {NoStop}%
\bibitem [{\citenamefont {Herbschleb}\ \emph {et~al.}(2022)\citenamefont
  {Herbschleb}, \citenamefont {Ohki}, \citenamefont {Morita}, \citenamefont
  {Yoshii}, \citenamefont {Kato}, \citenamefont {Makino}, \citenamefont
  {Yamasaki},\ and\ \citenamefont
  {Mizuochi}}]{herbschlebLowFrequencyQuantumSensing2022}%
  \BibitemOpen
  \bibfield  {author} {\bibinfo {author} {\bibfnamefont {E.}~\bibnamefont
  {Herbschleb}}, \bibinfo {author} {\bibfnamefont {I.}~\bibnamefont {Ohki}},
  \bibinfo {author} {\bibfnamefont {K.}~\bibnamefont {Morita}}, \bibinfo
  {author} {\bibfnamefont {Y.}~\bibnamefont {Yoshii}}, \bibinfo {author}
  {\bibfnamefont {H.}~\bibnamefont {Kato}}, \bibinfo {author} {\bibfnamefont
  {T.}~\bibnamefont {Makino}}, \bibinfo {author} {\bibfnamefont
  {S.}~\bibnamefont {Yamasaki}},\ and\ \bibinfo {author} {\bibfnamefont
  {N.}~\bibnamefont {Mizuochi}},\ }\bibfield  {title} {\bibinfo {title}
  {Low-{{Frequency Quantum Sensing}}},\ }\href
  {https://doi.org/10.1103/PhysRevApplied.18.034058} {\bibfield  {journal}
  {\bibinfo  {journal} {Physical Review Applied}\ }\textbf {\bibinfo {volume}
  {18}},\ \bibinfo {pages} {034058} (\bibinfo {year} {2022})}\BibitemShut
  {NoStop}%
\bibitem [{\citenamefont {Lekavicius}\ \emph {et~al.}(2023)\citenamefont
  {Lekavicius}, \citenamefont {Carter}, \citenamefont {Pennachio},
  \citenamefont {White}, \citenamefont {Hajzus}, \citenamefont {Purdy},
  \citenamefont {Gaskill}, \citenamefont {Yeats},\ and\ \citenamefont
  {{Myers-Ward}}}]{lekaviciusMagnetometryBasedSiliconVacancy2023}%
  \BibitemOpen
  \bibfield  {author} {\bibinfo {author} {\bibfnamefont {I.}~\bibnamefont
  {Lekavicius}}, \bibinfo {author} {\bibfnamefont {S.}~\bibnamefont {Carter}},
  \bibinfo {author} {\bibfnamefont {D.}~\bibnamefont {Pennachio}}, \bibinfo
  {author} {\bibfnamefont {S.}~\bibnamefont {White}}, \bibinfo {author}
  {\bibfnamefont {J.}~\bibnamefont {Hajzus}}, \bibinfo {author} {\bibfnamefont
  {A.}~\bibnamefont {Purdy}}, \bibinfo {author} {\bibfnamefont
  {D.}~\bibnamefont {Gaskill}}, \bibinfo {author} {\bibfnamefont
  {A.}~\bibnamefont {Yeats}},\ and\ \bibinfo {author} {\bibfnamefont
  {R.}~\bibnamefont {{Myers-Ward}}},\ }\bibfield  {title} {\bibinfo {title}
  {Magnetometry {{Based}} on {{Silicon-Vacancy Centers}} in {{Isotopically
  Purified 4H-SiC}}},\ }\href
  {https://doi.org/10.1103/PhysRevApplied.19.044086} {\bibfield  {journal}
  {\bibinfo  {journal} {Physical Review Applied}\ }\textbf {\bibinfo {volume}
  {19}},\ \bibinfo {pages} {044086} (\bibinfo {year} {2023})}\BibitemShut
  {NoStop}%
\bibitem [{\citenamefont {Kraus}\ \emph {et~al.}(2014)\citenamefont {Kraus},
  \citenamefont {Soltamov}, \citenamefont {Fuchs}, \citenamefont {Simin},
  \citenamefont {Sperlich}, \citenamefont {Baranov}, \citenamefont {Astakhov},\
  and\ \citenamefont {Dyakonov}}]{krausMagneticFieldTemperature2014}%
  \BibitemOpen
  \bibfield  {author} {\bibinfo {author} {\bibfnamefont {H.}~\bibnamefont
  {Kraus}}, \bibinfo {author} {\bibfnamefont {V.~A.}\ \bibnamefont {Soltamov}},
  \bibinfo {author} {\bibfnamefont {F.}~\bibnamefont {Fuchs}}, \bibinfo
  {author} {\bibfnamefont {D.}~\bibnamefont {Simin}}, \bibinfo {author}
  {\bibfnamefont {A.}~\bibnamefont {Sperlich}}, \bibinfo {author}
  {\bibfnamefont {P.~G.}\ \bibnamefont {Baranov}}, \bibinfo {author}
  {\bibfnamefont {G.~V.}\ \bibnamefont {Astakhov}},\ and\ \bibinfo {author}
  {\bibfnamefont {V.}~\bibnamefont {Dyakonov}},\ }\bibfield  {title} {\bibinfo
  {title} {Magnetic field and temperature sensing with atomic-scale spin
  defects in silicon carbide},\ }\href {https://doi.org/10.1038/srep05303}
  {\bibfield  {journal} {\bibinfo  {journal} {Scientific Reports}\ }\textbf
  {\bibinfo {volume} {4}},\ \bibinfo {pages} {5303} (\bibinfo {year}
  {2014})}\BibitemShut {NoStop}%
\bibitem [{\citenamefont {Abraham}\ \emph {et~al.}(2021)\citenamefont
  {Abraham}, \citenamefont {Gutgsell}, \citenamefont {Todorovski},
  \citenamefont {Sperling}, \citenamefont {Epstein}, \citenamefont
  {{Tien-Street}}, \citenamefont {Sweeney}, \citenamefont {Wathen},
  \citenamefont {Pogue}, \citenamefont {Brereton}, \citenamefont {McQueen},
  \citenamefont {Frey}, \citenamefont {Clader},\ and\ \citenamefont
  {Osiander}}]{abrahamNanoteslaMagnetometrySilicon2021}%
  \BibitemOpen
  \bibfield  {author} {\bibinfo {author} {\bibfnamefont {J.~B.~S.}\
  \bibnamefont {Abraham}}, \bibinfo {author} {\bibfnamefont {C.}~\bibnamefont
  {Gutgsell}}, \bibinfo {author} {\bibfnamefont {D.}~\bibnamefont
  {Todorovski}}, \bibinfo {author} {\bibfnamefont {S.}~\bibnamefont
  {Sperling}}, \bibinfo {author} {\bibfnamefont {J.~E.}\ \bibnamefont
  {Epstein}}, \bibinfo {author} {\bibfnamefont {B.~S.}\ \bibnamefont
  {{Tien-Street}}}, \bibinfo {author} {\bibfnamefont {T.~M.}\ \bibnamefont
  {Sweeney}}, \bibinfo {author} {\bibfnamefont {J.~J.}\ \bibnamefont {Wathen}},
  \bibinfo {author} {\bibfnamefont {E.~A.}\ \bibnamefont {Pogue}}, \bibinfo
  {author} {\bibfnamefont {P.~G.}\ \bibnamefont {Brereton}}, \bibinfo {author}
  {\bibfnamefont {T.~M.}\ \bibnamefont {McQueen}}, \bibinfo {author}
  {\bibfnamefont {W.}~\bibnamefont {Frey}}, \bibinfo {author} {\bibfnamefont
  {B.~D.}\ \bibnamefont {Clader}},\ and\ \bibinfo {author} {\bibfnamefont
  {R.}~\bibnamefont {Osiander}},\ }\bibfield  {title} {\bibinfo {title}
  {Nanotesla {{Magnetometry}} with the {{Silicon Vacancy}} in {{Silicon
  Carbide}}},\ }\href {https://doi.org/10.1103/PhysRevApplied.15.064022}
  {\bibfield  {journal} {\bibinfo  {journal} {Physical Review Applied}\
  }\textbf {\bibinfo {volume} {15}},\ \bibinfo {pages} {064022} (\bibinfo
  {year} {2021})}\BibitemShut {NoStop}%
\bibitem [{\citenamefont {Naydenov}\ \emph {et~al.}(2011)\citenamefont
  {Naydenov}, \citenamefont {Dolde}, \citenamefont {Hall}, \citenamefont
  {Shin}, \citenamefont {Fedder}, \citenamefont {Hollenberg}, \citenamefont
  {Jelezko},\ and\ \citenamefont
  {Wrachtrup}}]{naydenovDynamicalDecouplingSingleelectron2011}%
  \BibitemOpen
  \bibfield  {author} {\bibinfo {author} {\bibfnamefont {B.}~\bibnamefont
  {Naydenov}}, \bibinfo {author} {\bibfnamefont {F.}~\bibnamefont {Dolde}},
  \bibinfo {author} {\bibfnamefont {L.~T.}\ \bibnamefont {Hall}}, \bibinfo
  {author} {\bibfnamefont {C.}~\bibnamefont {Shin}}, \bibinfo {author}
  {\bibfnamefont {H.}~\bibnamefont {Fedder}}, \bibinfo {author} {\bibfnamefont
  {L.~C.~L.}\ \bibnamefont {Hollenberg}}, \bibinfo {author} {\bibfnamefont
  {F.}~\bibnamefont {Jelezko}},\ and\ \bibinfo {author} {\bibfnamefont
  {J.}~\bibnamefont {Wrachtrup}},\ }\bibfield  {title} {\bibinfo {title}
  {Dynamical decoupling of a single-electron spin at room temperature},\ }\href
  {https://doi.org/10.1103/PhysRevB.83.081201} {\bibfield  {journal} {\bibinfo
  {journal} {Physical Review B}\ }\textbf {\bibinfo {volume} {83}},\ \bibinfo
  {pages} {081201} (\bibinfo {year} {2011})}\BibitemShut {NoStop}%
\bibitem [{\citenamefont {{de Lange}}\ \emph {et~al.}(2011)\citenamefont {{de
  Lange}}, \citenamefont {Rist{\`e}}, \citenamefont {Dobrovitski},\ and\
  \citenamefont {Hanson}}]{delangeSingleSpinMagnetometryMultipulse2011}%
  \BibitemOpen
  \bibfield  {author} {\bibinfo {author} {\bibfnamefont {G.}~\bibnamefont {{de
  Lange}}}, \bibinfo {author} {\bibfnamefont {D.}~\bibnamefont {Rist{\`e}}},
  \bibinfo {author} {\bibfnamefont {V.~V.}\ \bibnamefont {Dobrovitski}},\ and\
  \bibinfo {author} {\bibfnamefont {R.}~\bibnamefont {Hanson}},\ }\bibfield
  {title} {\bibinfo {title} {Single-{{Spin Magnetometry}} with {{Multipulse
  Sensing Sequences}}},\ }\href
  {https://doi.org/10.1103/PhysRevLett.106.080802} {\bibfield  {journal}
  {\bibinfo  {journal} {Physical Review Letters}\ }\textbf {\bibinfo {volume}
  {106}},\ \bibinfo {pages} {080802} (\bibinfo {year} {2011})}\BibitemShut
  {NoStop}%
\bibitem [{\citenamefont {Lekavicius}\ \emph {et~al.}(2022)\citenamefont
  {Lekavicius}, \citenamefont {{Myers-Ward}}, \citenamefont {Pennachio},
  \citenamefont {Hajzus}, \citenamefont {Gaskill}, \citenamefont {Purdy},
  \citenamefont {Yeats}, \citenamefont {Brereton}, \citenamefont {Glaser},
  \citenamefont {Reinecke},\ and\ \citenamefont
  {Carter}}]{lekaviciusOrdersMagnitudeImprovement2022}%
  \BibitemOpen
  \bibfield  {author} {\bibinfo {author} {\bibfnamefont {I.}~\bibnamefont
  {Lekavicius}}, \bibinfo {author} {\bibfnamefont {R.}~\bibnamefont
  {{Myers-Ward}}}, \bibinfo {author} {\bibfnamefont {D.}~\bibnamefont
  {Pennachio}}, \bibinfo {author} {\bibfnamefont {J.}~\bibnamefont {Hajzus}},
  \bibinfo {author} {\bibfnamefont {D.}~\bibnamefont {Gaskill}}, \bibinfo
  {author} {\bibfnamefont {A.}~\bibnamefont {Purdy}}, \bibinfo {author}
  {\bibfnamefont {A.}~\bibnamefont {Yeats}}, \bibinfo {author} {\bibfnamefont
  {P.}~\bibnamefont {Brereton}}, \bibinfo {author} {\bibfnamefont
  {E.}~\bibnamefont {Glaser}}, \bibinfo {author} {\bibfnamefont
  {T.}~\bibnamefont {Reinecke}},\ and\ \bibinfo {author} {\bibfnamefont
  {S.}~\bibnamefont {Carter}},\ }\bibfield  {title} {\bibinfo {title} {Orders
  of {{Magnitude Improvement}} in {{Coherence}} of {{Silicon-Vacancy
  Ensembles}} in {{Isotopically Purified 4H-SiC}}},\ }\href
  {https://doi.org/10.1103/PRXQuantum.3.010343} {\bibfield  {journal} {\bibinfo
   {journal} {PRX Quantum}\ }\textbf {\bibinfo {volume} {3}},\ \bibinfo {pages}
  {010343} (\bibinfo {year} {2022})}\BibitemShut {NoStop}%
\bibitem [{\citenamefont
  {Tahara}(2023)}]{taharaMAHOSMeasurementAutomation2023}%
  \BibitemOpen
  \bibfield  {author} {\bibinfo {author} {\bibfnamefont {K.}~\bibnamefont
  {Tahara}},\ }\bibfield  {title} {\bibinfo {title} {{{MAHOS}}: {{Measurement
  Automation Handling}} and {{Orchestration System}}},\ }\href
  {https://doi.org/10.21105/joss.05938} {\bibfield  {journal} {\bibinfo
  {journal} {Journal of Open Source Software}\ }\textbf {\bibinfo {volume}
  {8}},\ \bibinfo {pages} {5938} (\bibinfo {year} {2023})}\BibitemShut
  {NoStop}%
\end{thebibliography}%

\end{document}